\documentclass[a4paper,12pt]{article}
\pdfoutput=1
\usepackage{epsfig}
\usepackage{amssymb,subfigure}
\usepackage{amsfonts}
\usepackage{amsmath}
\usepackage{euscript}
\usepackage{verbatim}
\usepackage{latexsym}
\usepackage[compat=1.0.0]{tikz-feynman}
\usepackage{graphicx,color}
\usepackage{caption}
\usepackage{float}
\usepackage{slashed}
\usepackage{graphicx}
\graphicspath{ {images/} }
\usepackage{ytableau}
\usepackage{mathtools}

\usepackage[colorinlistoftodos]{todonotes}
\usepackage[colorlinks=true, allcolors=blue]{hyperref}

\usepackage{wrapfig}
	\usepackage[T1]{fontenc}
	\usepackage{tikz}
	\usetikzlibrary{decorations.pathmorphing}
\usepackage{tikz}
\usetikzlibrary{shapes.geometric, arrows,patterns,snakes}
\tikzstyle{ellip} = [ellipse, minimum width=3cm, minimum height=1cm,text centered, draw=black]

\newskip\humongous \humongous=0pt plus 1000pt minus 1000pt

\newif\ifdtup

\allowdisplaybreaks[1]

\catcode`\@=11

\@addtoreset{equation}{section}

\def\@normalsize{\@setsize\normalsize{15pt}\xiipt\@xiipt
\abovedisplayskip 14pt plus3pt minus3pt%
\belowdisplayskip \abovedisplayskip
\abovedisplayshortskip \z@ plus3pt%
\belowdisplayshortskip 7pt plus3.5pt minus0pt}

\def\small{\@setsize\small{13.6pt}\xipt\@xipt
\abovedisplayskip 13pt plus3pt minus3pt%
\belowdisplayskip \abovedisplayskip
\abovedisplayshortskip \z@ plus3pt%
\belowdisplayshortskip 7pt plus3.5pt minus0pt
\def\@listi{\parsep 4.5pt plus 2pt minus 1pt
     \itemsep \parsep
     \topsep 9pt plus 3pt minus 3pt}}

\relax

\catcode`@=12

\catcode`\@=11

\def\section{\@startsection{section}{1}{\z@}{3.5ex plus 1ex minus
   .2ex}{2.3ex plus .2ex}{\large\bf}}

\def\SymBoxes#1#2#3#4{\newdimen\un@t \un@t#3%
\raisebox{#1}{\rule{#2\un@t}{#4}\hskip-#2\un@t
\@tempdimb\un@t \advance\@tempdimb by-#4\@tempcntb#2\relax%
\@whilenum{\@tempcntb>0}\do{
\rule{#4}{\un@t}\hskip\@tempdimb \advance\@tempcntb by\m@ne}%
\hskip-#2\un@t \rule[\un@t]{#2\un@t}{#4}%
\rule[\un@t]{#4}{#4}\hskip-#4
\rule{#4}{\un@t}}\hskip-#4}                

\begin{document}

\newcommand{\beq}{\begin{equation}}
\newcommand{\eeq}{\end{equation}}
\newcommand{\bea}{\begin{eqnarray}}
\newcommand{\eea}{\end{eqnarray}}
\newcommand{\beas}{\begin{eqnarray*}}
\newcommand{\eeas}{\end{eqnarray*}}
\newcommand{\defi}{\stackrel{\rm def}{=}}
\newcommand{\non}{\nonumber}
\newcommand{\bquo}{\begin{quote}}
\newcommand{\enqu}{\end{quote}}
\renewcommand{\(}{\begin{equation}}
\renewcommand{\)}{\end{equation}}
\def \eqn#1#2{\begin{equation}#2\label{#1}\end{equation}}
\def\IZ{{\mathbb Z}}
\def\IR{{\mathbb R}}
\def\IC{{\mathbb C}}
\def\IQ{{\mathbb Q}}
\def\de{\partial}
\def\Tr{ \hbox{\rm Tr}}
\def\H{ \hbox{\rm H}}
\def\HE{ \hbox{$\rm H^{even}$}}
\def\HO{ \hbox{$\rm H^{odd}$}}
\def\K{ \hbox{\rm K}}
\def\Im{ \hbox{\rm Im}}
\def\Ker{ \hbox{\rm Ker}}
\def\const{\hbox {\rm const.}}
\def\o{\over}
\def\im{\hbox{\rm Im}}
\def\re{\hbox{\rm Re}}
\def\bra{\langle}\def\ket{\rangle}
\def\Arg{\hbox {\rm Arg}}
\def\Re{\hbox {\rm Re}}
\def\Im{\hbox {\rm Im}}
\def\exo{\hbox {\rm exp}}
\def\diag{\hbox{\rm diag}}
\def\longvert{{\rule[-2mm]{0.1mm}{7mm}}\,}
\def\a{\alpha}
\def\dag{{}^{\dagger}}
\def\tq{{\widetilde q}}
\def\p{{}^{\prime}}
\def\W{W}
\def\N{{\cal N}}
\def\hsp{,\hspace{.7cm}}

\def\br{\nonumber\\}
\def\IZ{{\mathbb Z}}
\def\IR{{\mathbb R}}
\def\IC{{\mathbb C}}
\def\IQ{{\mathbb Q}}
\def\IP{{\mathbb P}}
\def \eqn#1#2{\begin{equation}#2\label{#1}\end{equation}}

\newcommand{\sgm}[1]{\sigma_{#1}}
\newcommand{\idd}{\mathbf{1}}

\newcommand{\C}{\ensuremath{\mathbb C}}
\newcommand{\Z}{\ensuremath{\mathbb Z}}
\newcommand{\R}{\ensuremath{\mathbb R}}
\newcommand{\rp}{\ensuremath{\mathbb {RP}}}
\newcommand{\cp}{\ensuremath{\mathbb {CP}}}
\newcommand{\vac}{\ensuremath{|0\rangle}}
\newcommand{\vact}{\ensuremath{|00\rangle}                    }
\newcommand{\oc}{\ensuremath{\overline{c}}}
\begin{titlepage}
\begin{flushright}
CHEP XXXXX
\end{flushright}
\bigskip
\def\thefootnote{\fnsymbol{footnote}}

\begin{center}
{\Large
{\bf Massive Scattering Amplitudes in Six Dimensions \\
\vspace{0.2in}
}
}
\end{center}

\bigskip
\begin{center}
{\large  Rishabh JHA\footnote{\texttt{rishabh.jha77@gmail.com}}, Chethan KRISHNAN$^a$\footnote{\texttt{chethan.krishnan@gmail.com}}, K. V. Pavan KUMAR$^a$\footnote{\texttt{kumar.pavan56@gmail.com}} \vspace{0.15in} \\ }
\vspace{0.1in}

\end{center}

\renewcommand{\thefootnote}{\arabic{footnote}}

\begin{center}
$^a$ {Center for High Energy Physics,\\
Indian Institute of Science, Bangalore 560012, India}

\end{center}

\noindent
\begin{center} {\bf Abstract} \end{center}
We show that a natural spinor-helicity formalism that can describe massive scattering amplitudes exists in $D=6$ dimensions. This is arranged by having helicity spinors carry an index in the Dirac spinor {\bf 4} of the massive little group, $SO(5) \sim Sp(4)$. In the high energy limit, two separate kinds of massless helicity spinors emerge as required for consistency with arXiv:0902.0981, with indices in the two $SU(2)$'s of the massless little group $SO(4)$. The tensors of ${\bf 4}$ lead to particles with arbitrary spin, and using these and demanding consistent factorization, we can fix $3-$ and $4-$point tree amplitudes of arbitrary masses and spins: we provide examples. We discuss the high energy limit of scattering amplitudes and the Higgs mechanism in this language, and make some preliminary observations about massive BCFW recursion.

\vspace{1.6 cm}
\vfill

\end{titlepage}

\setcounter{page}{2}
\tableofcontents

\setcounter{footnote}{0}


\section{Introduction}

The philosophy that one can fix a theory based largely on consistency conditions (like unitarity, locality, Lorentz invariance) is an old one and its incarnation in S-matrix bootstrap is sometimes considered to be the roots from which string theory arose. A closely related idea that has lead to substantial progress in the last decade is to use on-shell methods (without direct reliance on local quantum fields) to compute scattering amplitudes. See \cite{TASI} for a review. The foundational observation here is that the basic objects in a quantum field theory can be viewed as particles and not fields, and the former transform in representations of the little group. Quantum fields (which are introduced as a tool for ensuring manifest locality) on the other hand transform as tensors of the Lorentz group, and it has become increasingly plausible in recent years that working with little group covariant objects (spinor helicity variables) might be a more natural and simpler way to construct scattering amplitudes, since they deal directly with particles. 

Typically, most work on scattering amplitudes is in the context of massless particles. This is partly because quantum field theories are typically more complicated in the context of massless particles (they often lead to gauge theories), and therefore the simplifications of the spinor helicity formalism are most apparent for massless particles. However, if our goals are truly ambitious, and we are trying to {\em derive} a UV completion like string theory from our general expectations about scattering amplitudes, then it is evident that we are likely to require the ability to deal with massive particle as well. In such a set up, scattering amplitudes of massive particles should be understandable in terms of IR deformations of massless scattering amplitudes. This type of a formalism for dealing with massive scattering amplitudes has indeed been developed recently in four dimensions by \cite{Nima} via an extension of the usual massless helicity spinors. In this paper, we will generalize this approach to six dimensions. 

Even though the general philosophy of using little group covariant quantities to describe scattering applies across dimensions, it has turned out that its usefulness is immediate only in certain dimensions: $D=3,4,6$, see \cite{Elvang}. This is because to take full advantage of helicity covariance one needs to work with variables that are manifestly on-shell, and explicitly solving on-shell constraints in a useful way is easiest in these dimensions\footnote{These specific dimensions have connections to the possible division algebras. See \cite{all dimensions} for a discussion in general dimensions.}. Among these, the massive case for four has already been done as we noted, three is somewhat trivial (though possibly still quite interesting for various purposes), and therefore we turn to six to see whether we can construct a massive spinor helicity formalism. We find that we can.   

\section{Spinor Helicity in Six Dimensions}

 
Before proceeding further, we note that any particle can be described by specifying its transformation properties under the little group. In six dimensions, the Lorentz group is $SO(5,1)\sim SU(4)$ and the little group is $SO(4)\sim SU(2)\times SU(2)$ for massless particles and $SO(5)\sim Sp(4)$ for massive particles. We denote the $SU(4)$ indices using $A,B,\ldots $ and the $Sp(4)$ indices are denoted by $I,J,\ldots$ whereas the undotted ($a,b,\ldots$) and dotted indices ($\dot{a},\dot{b},\ldots$) correspond to the two $SU(2)$ groups.

\subsection{Spinor Helicity for Massless Particles in $6D$}

We begin by reviewing the spinor helicity formalism for massless particles following \cite{Cheung}. We start by decomposing the momentum as follows:
\begin{align}
\label{lambdatilde}
p_{AB}&\equiv p_{\mu} \Sigma ^{\mu}_{AB}=\tilde{\lambda }_{A\dot{a}} \tilde{\lambda }_{B\dot{b}}\epsilon ^{\dot{a}\dot{b}}\\
\label{lambda}
p^{AB}&\equiv p_{\mu} \bar{\Sigma} ^{\mu AB}=\lambda ^{A{a}} \lambda ^{B{b}}\epsilon _{{a}{b}}
\end{align}
The subscript and superscript Lorentz indices $\{A,B\}$ correspond to fundamental and anti-fundamental representations of $SU(4)$. $\Sigma $ and $\bar{\Sigma} $ correspond to gamma matrices in six dimensions and their explicit forms are given in the appendix. $\epsilon_{{a}{b}} $ are used to rise and lower the little group ($SU(2)\times SU(2)$ for massless particles) indices and we work with a convention where  $\epsilon_{21}=\epsilon^{12}=1$ for both the $SU(2)$ groups. Note that $p_{AB}$ and $p^{AB}$ are $4\times 4$  matrices of rank two. Also, $p_{AB}$ and $p^{AB}$ are related as follows:
\begin{align}
p^{AB}&=\frac{1}{2} \epsilon^{ABCD} p_{CD}
\end{align}

 Further, we demand that $\lambda^{Aa }$ and $\tilde{\lambda }_{A\dot{a}}$ satisfy the Dirac equation i.e., we need that $\lambda^{Aa }$ and $\tilde{\lambda }_{A\dot{a}}$  solve the following equations:
\begin{align}
\label{massless Dirac}
p_{AB} \lambda^{Ba }=0; ~~~ p^{AB}\tilde{\lambda }_{A\dot{a}}=0
\end{align}
From the explicit forms of $p_{AB}$ and $p^{AB}$, we can see that $\lambda^{Aa }$ and $\tilde{\lambda }_{A\dot{a}}$  are solutions the above equations only if the following relation holds:
\begin{align}
\lambda ^{A{a}}\tilde{\lambda }_{A\dot{b}}&=0
\end{align} 
The explicit forms of $\lambda  $ and $\tilde{\lambda } $ are given in the appendix. 

Similar to the case of $4D$, we can also introduce the notation of angle and square brackets here. We use the following notation:
\begin{align}
	\lambda ^{A{a}}=|p^a\rangle ^A; ~~\tilde{\lambda }_{A\dot{b}}=|p_{\dot{b}}]_A
\end{align}
Equations \eqref{lambdatilde} and \eqref{lambda} can be written in this notation as follows:
\begin{align}
p_{AB}=|p_{\dot{a}}]_A |p_{\dot{b}}]_B \epsilon^{\dot{a}\dot{b}}; ~~ p^{AB}=|p^a\rangle ^A |p^b\rangle ^B \epsilon_{{a}{b}}
\end{align} 

Before going on to the massive case, let us see how the degrees of freedom counting works in \eqref{lambdatilde}.  The momentum has five independent\footnote{Our counting is for complex momenta.} parameters. $\tilde{\lambda }_{A\dot{a}}$ has eight independent components initially.  But note that the RHS is invariant under one of the $SU(2)$'s of the little group . Since $SU(2)$ has three free complex parameters, we can use them to fix three of the components of $\tilde{\lambda }_{A\dot{a}}$ leaving us with a total of five independent components. A similar counting works for the $\lambda^{Aa }$ in \eqref{lambda}.

\subsection{Spinor Helicity for Massive Particles in $6D$}

Now, we turn to the massive case i.e., we consider particles whose momenta satisfy\footnote{Note that we are working with mostly negative signature.} $p_{\mu}p^{\mu}=m^2$ where $m$ is the mass of the particle. Here, we decompose the momentum as follows:
\begin{align}
\label{lambdatilde-massive}
p_{AB}&=\tilde{\lambda }_{AI}\tilde{\lambda }_{BJ}J^{IJ}\equiv |p_I]_A |p_J]_B J^{IJ}\\
\label{lambda-massive}
p^{AB}&=\lambda^{AI}\lambda^{BJ} J_{IJ}\equiv |p^I\rangle ^A|p^J\rangle ^B J_{IJ}
\end{align} 
where $\{I,J\}$ are $Sp(4)$ fundamental/ anti-fundamental representation indices and $J$ is an invariant matrix of $Sp(4)$ and its explicit form is given in the appendix\footnote{The explicit form of $J$ is basis dependent but our construction goes through if we do a symplectic rotation appropriately on everything.}. Using the matrix $J$, we can raise and lower the $Sp(4)$ indices. Also, starting from the following property of symplectic matrices:
\begin{align}
M^T J M=J
\end{align} 
where $M\in Sp(4)$, we can show that $p_{AB}$ and $p^{AB}$ are invariant under $Sp(4)$ transformations of $\lambda $ and $\tilde{\lambda }$. Further, demanding that $\lambda $ and $\tilde{\lambda }$ satisfy the Dirac equation\footnote{The Dirac equation in the massive case is given as follows: $p_{AB} \lambda^{A I} = m \tilde{\lambda}_B^I, ~~ p^{AB} \tilde{\lambda}_A^I = m \lambda^{B I}$} gives us the following condition:
\begin{align}
\label{lambdas relation}
\lambda^{AI }\tilde{\lambda }_{A}^J=-m J^{IJ}
\end{align}

Let us now explain how the counting works in the massive case. Consider equation \eqref{lambdatilde-massive}. The momentum has six independent parameters where mass is one of them. $\tilde{\lambda } $ is a $4\times 4$ matrix and hence there are 16 independent parameters. But, as explained above, $p_{AB}$ is invariant under $Sp(4)$. Hence, we can use ten free parameters of $Sp(4)$ to fix ten components of $\tilde{\lambda } $ leaving us with six degrees\footnote{Note that in 6d, we can find $\lambda $'s such that their product gives momentum and they satisfy Dirac equation simultaneously. This can be checked using the explicit forms of $\lambda $'s we constructed in the appendix. Hence the equation \eqref{lambdas relation} does not impose any more constraints on $\lambda $'s.} of freedom as desired. A similar counting works for $\lambda $ as well. 

\subsection{The High-Energy Limit}
\label{high-energy-limit}

We conclude this section by commenting on how to connect our formalism for massive particles with that of the spinor helicity formalism of massless particles. Note that this will also be useful later in the paper where we take the high-energy limit of various massive scattering amplitudes and match it with the results of massless amplitudes presented in \cite{Cheung}. 

We start by writing down\footnote{Algebraically, we are expanding a $4\times 4 $ matrix in the basis of $4\times 2 $ matrices. } the massive helicity spinors in $6D$ in terms of their massless counterparts as follows:
\begin{align}
\lambda^{A I} &= \sqrt{\frac{E+p}{2 p}} \lambda^A_a e^{+I a} + \sqrt{\frac{E-p}{2p}} \eta^{A \dot{a}} e^{-I}_{\dot{a}}\\
\tilde{\lambda}_A^I &= \sqrt{\frac{E+p}{2 p}} \tilde{\lambda}_A^{\dot{a}} e^{-I}_{\dot{a}} + \sqrt{\frac{E-p}{2 p}} \tilde{\eta}_{A a} e^{+I a}
\end{align}
where $e$'s are the basis vectors and their explicit forms are given in the appendix\footnote{These $e$'s are a type of projection matrices. See equation (2.20) in \cite{Nima} for a similar construction in 4d where things are a bit simpler and therefore are more intuitive.}. $p$ is the magnitude of the spatial part of the momentum and $E$ is the energy of the particle i.e., $p^0=E$. $\{\lambda _a, \tilde{\lambda }^{\dot{a}}\}$ and $\{\eta^{\dot{a}}, \tilde{\eta }_{a}\}$ are massless helicity spinors that satisfy the massless Dirac equation \eqref{massless Dirac} where we compute $p_{AB}/ p^{AB}$ with $E=+p$ and $E=-p$ respectively. These spinors satisfy the relations $p^{AB} = \lambda^{AI} \lambda^{B}_I$ and $p_{AB} = \tilde{\lambda}_{AI} \tilde{\lambda}_{B}^I$. Furthermore the above expansions are useful in obtaining explicit forms of $\lambda ^I$ and $\tilde{\lambda }^J$ because it is easier to solve for the massless helicity spinors using \eqref{massless Dirac} as they are decoupled equations as opposed to massive Dirac equation \eqref{massiveDirac} which couples $\lambda^I$ and $\tilde{\lambda }^J$. We use this strategy to find the explicit forms of $\lambda ^I$ and $\tilde{\lambda }^J$ in the appendix.

The high-energy limit is taken in the later sections as discussed henceforth. Starting from the massive scattering amplitudes, we use the above expansions for $\lambda $ and $\tilde{\lambda } $ to express the massive amplitudes in terms of massless amplitudes as follows:
\begin{align}
{\cal M}^{I_1\ldots I_N}=\sum_i \left({e^{_+}}^i {e^{_-}}^{N-i}\right) ^{I_1\ldots I_N; ~a_1\ldots a_i}_{\dot{a}_{i+1}\ldots \dot{a}_N} {\cal M}^{\dot{a}_{i+1}\ldots \dot{a}_N}_{a_1\ldots a_i} 
\end{align}
By writing ${e^{_+}}^i {e^{_-}}^{N-i}$, we mean the following. For any $i$, we have $i$ factors of $e^+$'s and $(N-i)$ factors of $e^-$'s. But for convenience in notation above, we have collected all the $(I,a,\dot{a})$ indices separately outside. Then the high-energy limit is taken for each term separately i.e.,
\begin{align}
\text{HE limit of}~{e^{_+}}^i {e^{_-}}^{N-i}~\text{component}&=\lim\limits_{m\rightarrow 0} {\cal M}^{\dot{a}_{i+1}\ldots \dot{a}_N}_{a_1\ldots a_i} 
\end{align}
An example on how to take the high-energy limit is given in the section after the discussion of three point and four point functions.

\section{Three Point Scattering Amplitudes}
\label{3-points}
In this section, we present a strategy to construct all the possible three point amplitudes in six dimensions where at least one of the particles is massive.

To construct these amplitudes, we use the fact that the scattering amplitude is Lorentz invariant and little group covariant with respect to each of the particles. That is,
\begin{align}
{\cal M}\left(\{|1\rangle ,|1]\},\ldots \{S|i\rangle ,S|i]\} \ldots \right)= S {\cal M}\left(\{|1\rangle ,|1]\},\ldots \{|i\rangle ,|i]\} \ldots \right)
\end{align} 
where $S$ is an arbitrary little group transformation i.e., $S\in Sp(4)$ for massive particles and $S\in SU(2)\times SU(2)$ for massless particles.

Before going ahead, we mention that various coupling constants that appear in the following are \textit{not} necessarily mass dimension zero. Their mass dimensions are to be fixed such that the entire amplitude has appropriate dimension. For example, in equations \eqref{1-massive} or \eqref{1-massive example}, the $g_{\alpha}$'s are assumed to have the same dimension. Note that this does \textit{not} lead to any loss of generality because we can include masses of particles in the definition of Lorentz tensors $M$ (see the next subsections for the definition of $M$) to soak up extra dimensions.

\subsection{$2$-Massless, $1$-Massive Particles}

Let us start the construction of three point amplitudes where only one of the particles is massive. We take the particles $1$ and $2$ to be massless and the third particle to be massive with mass $m_3$. Here we consider particles\footnote{Note that the rest of the discussion goes through even when 
these two particles have more than two little group indices each but to avoid clutter of notation, we work with this choice.} $1$ and $2$ to have one dotted and one undotted index each. We denote them as $1^{a \dot{a}}$ and $2^{b \dot{b}}$ respectively. The massive particle has $N$ $Sp(4)$ indices and transforms in some representation $R$ of the little group $Sp(4)$ and we denote it as  $3^{\{I_1, I_2,..., I_{N}\}_R}$. Since the scattering amplitude is covariant under the little group and invariant under the Lorentz group, we can  write\footnote{In principle, the three point functions can include inverse factors such as $\langle 1^a2^{\dot{b}}]^{-1}$. But since $p_1.p_2\neq 0$, we can always rewrite them in terms of $\langle 1^a2^{\dot{b}}]$. Hence we need not worry about such inverses in our formalism.} it as follows:
\begin{align}
\mathcal{M}_{3}^{a b \dot{a} \dot{b} \{I_1 I_2 \ldots I_{N}\}_R}=\sum _{\alpha}g_{\alpha}S_{\alpha}^{a b \dot{a} \dot{b} \{I_1 I_2 \ldots I_{N}\}_R}
\end{align}
where each $S_{\alpha}$ is of the form:
\begin{align}
\label{1-massive}
S_{\alpha}^{a b \dot{a} \dot{b} \{I_1 I_2 \ldots I_{N}\}_R}= ~\lambda_1^{A_1 a} \lambda_2^{B_1 b} \tilde{\lambda}_{1 A_2}^{\dot{a}} \tilde{\lambda}_{2 B_2}^{\dot{b}}\tilde{\lambda}_{3 C_1}^{ I_1}\tilde{\lambda}_{3 C_2}^{ I_2}\ldots \tilde{\lambda}_{3 C_N}^{ I_N}~ M^{A_2 B_2\{C_1 C_2 \ldots C_{N}\}_R}_{A_1 B_1 }
\end{align}
where we have used the massive Dirac equation to write the  above amplitude only in terms of $\tilde{\lambda}_{3 }$'s.  By writing this equation, we have reformulated our problem of constructing scattering amplitudes to constructing the tensor $M$ of the Lorentz group $SU(4)$.  Note that the available\footnote{Here, we did not include $p_{iAB}$ in the list of available tensors because of the following relation: $\frac{1}{2}\epsilon^{ABCD} p_{ iCD} = p_i^{AB}$.} $SU(4)$ tensors to construct $M$ are $\delta ^A_B$, $\epsilon_{ABCD} $, $\epsilon^{ABCD}$ and $\frac{1}{m_3}p_{1,2,3} ^{AB}$.  Also, we need to construct $M$ such that the resultant scattering amplitude transforms in the representation $R$ of $Sp(4)$ with respect to $I_1,\ldots I_N$. This can be achieved if we demand that the Lorentz tensor $M$ should  transform in the representation $R$ with respect to the indices $C_1,\ldots C_N$ and we have already incorporated this in the above equation. Summarizing, $M$ is sum of all the possible terms that can be constructed using the available tensors of $SU(4)$ that have the right transformation properties. Each of these terms in $M$ can have a different coupling constant. We denote it by $g_{\alpha}$ where $\alpha $ corresponds to different terms in the sum. Sum over $\alpha $ is taken in \eqref{1-massive} to indicate sum over various $M$.


Before giving an example where we write down the scattering amplitude explicitly, let us give a simple counting argument that rules out certain interactions. Consider a general case where there are $n$ number of superscript indices and $m$ number of subscript indices in $M$. To construct $M$, suppose that we need to use $\alpha$ number of $\delta^C_D$, $\beta$ number of $\epsilon_{ABCD}$, $\gamma$ number of $\epsilon^{ABCD}$ and $\eta$ number of $\frac{1}{m_3}p_{i}^{AB}$. Then the following relation should hold:
\begin{align}
\label{counting}
n-m&=4\gamma +2\eta -4\beta
\end{align}
Since all the parameters here are non-negative integers, if $(n-m)$ is not even, then we can not construct $M$. Also, as $m$ and $n$ are  related to the number of little group indices, this relation can also be used to see whether a scattering is possible among the given particles. For example, consider the equation \eqref{1-massive}. The values of $n$ and $m$ are $(N+2)$ and $2$ respectively. From the above relation \eqref{counting}, it is clear that we can not write down any scattering amplitude if $N$ is odd. For instance, this statement implies that we can not have an interaction with two massless vector particles and a massive half-spin particle.

Next, we give a simple example where we write down the scattering amplitude of two massless fermions and one massive vector particle explicitly. That is, the third particle has two massive little group indices $I_1$ and $I_2$ that are in antisymmetric representation\footnote{Note that from the usual antisymmetric combination, the trace part has to be subtracted to obtain the antisymmetric representation of $Sp(4)$. See \cite{Georgi} for example.} of $Sp(4)$.  So, the particles we consider in this example are $1^a$, $2^{\dot{b}}$  and $3^{[I_1I_2]}$. Remembering that we can always convert between $\lambda$ and $\tilde{\lambda}$ using the massive $6D$ Dirac equation (\ref{massiveDirac}), we get the following expression:
\begin{equation}
\label{1-massive example}
\mathcal{M}_{3}^{a \dot{b} [I_1 I_2]} = \sum_{\alpha}g_{\alpha}
 ~\lambda_1^{A a} \tilde{\lambda}_{2B}^{ \dot{b}} \tilde{\lambda}_{3C}^{I_1} \tilde{\lambda}_{3 D}^{I_2}  ~M^{B [C D]}_{A}
\end{equation}
where the possibilities of $M$ are:
\begin{align}
M^{B [C D]}_{A}&=\frac{1}{m_3}\left(\delta ^{C}_Ap_3^{DB}-\delta ^{D}_Ap_3^{CB}-\frac{1}{2}\delta_A^B p_3^{CD}\right); ~\frac{1}{m_3} \delta_A ^B\left(p_1^{CD}-\frac{1}{2}p_3^{CD}\right)
\end{align}
The last term in each of the above expressions is required to remove the trace term in the resulting amplitude. So, the three point scattering amplitude for these particles can be written  as:
\begin{align}
{\cal M}_3^{a\dot{b}[I_1I_2]}&=g_1 \left(\langle 1^a3^{I_1}] \langle 3^{I_2}2^{\dot{b}}]-(I_1\leftrightarrow I_2)-\frac{m_3}{2}J^{I_1I_2}\langle 1^a2^{\dot{b}}] \right)+\frac{g_2}{m_3}\langle 1^a2^{\dot{b}}]\left([3^{I_1}|1|3^{I_2}]-\frac{m_3^2}{2}J^{I_1I_2}\right)
\end{align}
From this expression, it is easy to see that $J_{I_1I_2}{\cal M}_3^{a\dot{b}[I_1I_2]}=0$ as desired. Further note that starting with the identity $p_i^{AB}\epsilon^{{a}{b}}=\lambda_{i }^{Aa}\lambda_{i }^{Aa}-(a\leftrightarrow b)$, we can show that both the terms in the above expression are the same. Hence, the amplitude of two massless fermions (transforming in fundamental and anti-fundamental representations of $Sp(4)$) and a massive vector can be written as:
\begin{align}
\label{amp-2massless}
{\cal M}_3^{a\dot{b}[I_1I_2]}&=g \left(\langle 1^a3^{I_1}] \langle 3^{I_2}2^{\dot{b}}]-(I_1\leftrightarrow I_2)-\frac{m_3}{2}J^{I_1I_2}\langle 1^a2^{\dot{b}}] \right)
\end{align}

Just to connect it to the usual Lagrangian formulation, we note that a part of the above amplitude can be obtained from the following piece of Lagrangian:
\begin{align}
{\cal L}\supset g\lambda _1^{a}A^{\mu}\partial _{\mu}\tilde{\lambda}_{2}^{\dot{b}}
\end{align}
where $g$ is coupling constant. Using Feynman rules and writing all the indices explicitly, we get the amplitude (apart from a numerical factor) as:
\begin{align}
{\cal A}^{a\dot{b}[I_1I_2]}&=g\lambda _1^{Aa}\tilde{\lambda}_{2A}^{\dot{b}}p_2^{\mu}\varepsilon_{3\mu}^{[I_1I_2]}
\end{align}
Using the corresponding expression in \eqref{massive polarization} for massive polarization $\varepsilon _3$, we obtain the amplitude in \eqref{amp-2massless}.

Note that the corresponding term in the Lagrangian is not minimal because of our choice of the fermions in this example. If we would have chosen\footnote{That is, we need to choose both the fermions to be chiral or anti-chiral i.e., we choose either $(\lambda_{1 }^{Aa},\lambda_{2 }^{Bb})$ or $(\tilde{\lambda }_{1A}^{\dot{a}},\tilde{\lambda }_{2B}^{\dot{b}})$.} both the fermions to be transforming in (anti-) fundamental representation of the Lorentz group $SU(4)$, then the corresponding amplitude could be obtained from $g\lambda _1^{a}\slashed{A}\lambda_{2}^{b}$ which is like a ``minimal'' coupling term. 

\subsection{$1$-Massless, $2$-Massive Particles}

Next, we consider the case where the particles $1$ and $2$ are massive and particle $3$ is massless. Here, there are two cases. One where $m_1 \neq m_2$ and $m_1 = m_2$. The latter case is subtle and is somewhat qualitatively different from the former. We start with the unequal masses case. 

\subsubsection{Unequal Mass}

Let particle $3$ be massless and has one dotted and one undotted index i.e., it is denoted\footnote{As before, our method works for any number of indices and to avoid clutter of notation, we restrict to just two indices.} as $3^{a \dot{a}}$. We assume that the particles $1$ and $2$ transform in representations $R_1$ and $R_2$ respectively of the massive little group $Sp(4)$ and we denote these particles as $1^{\{ I_1, I_2, \ldots ,I_{N_1} \}_{R_1}}$ and $2^{\{ J_1, J_2, \ldots, J_{N_2} \}_{R_2}}$. Following similar steps as in the previous subsection, the scattering amplitude can be written as follows:
\begin{align}
\label{2-massive-equal-mass}
\mathcal{M}_{3}^{\{ I_1, \ldots ,I_{N_1} \}_{R_1} \{ J_1, \ldots, J_{N_2} \}_{R_2} a \dot{a}  }&= \nonumber \\ 
\sum _{\alpha}g_{\alpha}& ~ \tilde{\lambda}_{1 A_1}^{ I_1} \ldots \tilde{\lambda}_{1 A_{N_1}}^{ I_{N_1}} \lambda_{2}^{B_1 J_1} \ldots \lambda_{2}^{ B_{N_2} J_{N_2}} \lambda_3^{C_1 a} \tilde{\lambda}_{3 C_2}^{\dot{a}} ~ M^{C_2\{A_1 \ldots A_{N_1}\}_{R_1}  }_{C_1\{B_1 \ldots B_{N_2}\}_{R_2} }
\end{align}
The list of available $SU(4)$ tensors to construct $M$ in this case are $\delta ^A_B$, $\epsilon_{ABCD} $, $\epsilon^{ABCD}$, $\frac{1}{g(m)}p_i ^{AB}$ where $g(m)$ is an order one polynomial in $m_{1,2}$ and is required for dimensional reasons.

Consider the following example where massive particles $1$ and $2$ are both spin $\frac{1}{2}$ and particle $3$ is spin $1$, then the scattering amplitude is given by:
\begin{equation}
\mathcal{M}_3^{I;J;  a \dot{a}} = \sum _{\alpha}g_{\alpha} ~ \tilde{\lambda}_{1 A}^{ I} \lambda_{2}^{B J} \lambda_3^{C_1 a} \tilde{\lambda}_{3 C_2}^{\dot{a}} ~ M^{A C_2 }_{B C_1}
\end{equation}
where there are two possibilities for $M$, namely $M = \delta^A_{C_1} \delta^{C_2}_{B}$ and $M = \frac{1}{r_1(m)r_2(m)} p_1^{A C_2} p_{2 B C_1}$ where both $r_1(m)$ and $r_2(m)$ have mass dimension $1$. Hence, the amplitude can be written as:
\begin{align}
\label{unequalmass}
\mathcal{M}_3^{I;J;  a \dot{a}}&=g_1\langle 3^a1^I] \langle 2^J3^{\dot{a}}]+g_2\langle 3^a2^J] \langle 1^I3^{\dot{a}}]
\end{align}
where we have absorbed a dimensionless function of $m$ into the final definition of $g_2$.

The above amplitude can be obtained by the following term in the Lagrangian:
\begin{align}
{\cal L}\supset g\bar{\psi}\sigma ^{\mu\nu}\psi F_{\mu \nu}
\end{align}
where $\sigma ^{\mu\nu}=\left[\gamma ^{\mu},\gamma ^{\nu}\right]$ and $F_{\mu\nu}=\partial_{[\mu}A_{\nu]}$ is the field strength. The Dirac fermion $\psi $ in this case is a doublet of the form $\left( \begin{array}{c} \lambda_{1} \\ \tilde{\lambda }_2 \end{array} \right)$. Using Feynman rules and $p_3^{AB}\varepsilon_{3BC}^{a\dot{a}}\sim \lambda _3^{Aa}\tilde{\lambda }_{3C}^{\dot{a}}$, we see that this term in the Lagrangian gives $g\left(\langle 3^a2^J] \langle 1^I3^{\dot{a}}]+\langle 3^a1^I] \langle 2^J3^{\dot{a}}]\right)$ which is a specific case of \eqref{unequalmass}.

\subsubsection{Equal Mass}
\label{x-factor}
In the equal mass case where $m_1=m_2=m$, from the momentum conservation, we see that $p_3.p_{1,2} = 0$. From this, we can show\footnote{Without loss of generality we could have assumed momentum of particle $1$ instead of particle $2$.} that $ p_{2AB} \lambda_3^{A a} \lambda_{3a}^{B} = \langle 3|p_2| 3\rangle = 0$ and that $p_2^{AB} \tilde{\lambda}_{3 A \dot{a}} \tilde{\lambda}_{3 B}^{\dot{a}} = [3|p_2|3] = 0$. Since particle $3$ is massless, we also have $\langle 3^a|3^{\dot{b}}] = 0$. These two equations imply that $p_2 |3\rangle \propto |3]$. The proportionality constant contains an additional information that is used to modify the vertex factors in amplitude calculations similar in line with the massive $4D$ case as in \cite{Nima}. This constant of proportionality is defined through either of the following equations:
\begin{align}
\label{x-defined}
p_{2AB} \lambda_3^{A a} = m x^{a \dot{a}} \tilde{\lambda}_{3 B \dot{a}} &\Rightarrow ~ x^{a \dot{a}} = \frac{\langle \zeta ^b | 2 | 3^a \rangle}{m \langle \zeta^b 3_{\dot{a}}]} \\
\label{x}
p_2^{AB} \tilde{\lambda}_{3 A \dot{a}} = m y_{\dot{a} a} \lambda_3^{B a}&\Rightarrow  ~y_{\dot{a} a} = \frac{[\zeta ^{\dot{b}}| 2 | 3_{\dot{a}}]}{[ \zeta ^{\dot{b}}3^a \rangle}
\end{align}
where $\zeta$ is a reference spinor. Note that $x,y$ are dimensionless but transform nontrivially under the massless little group $SU(2)\times SU(2)$. From the above two equations, we can show that $x^{a\dot{a}}y_{\dot{a}b}=\delta _b^a$. Also, we can show that $x^{a\dot{a}}=- y^{a\dot{a}}$ using the following identities\footnote{The explicit form of the inverse of $\langle p^aq_{\dot{b}}]$ is for example given in \cite{Plefka}.}:
\begin{align}
p_{AB}q^{BC}&=-\frac{1}{2}\left(\delta ^C_Ap^{GH}-\delta ^G_A p^{CH}+\delta ^H_Ap^{CG}\right) q_{GH}\nonumber \\
\langle p_aq^{\dot{b}}]^{-1}&=-\frac{\langle p^aq_{\dot{b}}]}{2p.q}; ~~ \langle p^a|q|p^b\rangle =-2\epsilon^{{a}{b}} (p.q) 
\end{align} 
Hence, we use only $x^{a\dot{a}}$ in the further calculations.

With the introduction of this $x$-factor, the three point amplitudes are different as compared to previous cases in the sense that the scattering amplitudes are now written as  polynomials in $\lambda$, $\tilde{\lambda}$ and $x$. Let us consider the first example where particles $1$ and $2$ are massive scalars with equal mass $m$ and particle $3$ is a massless vector with one dotted and one undotted index. Then the most general polynomial with proper transformation properties constructed out of $x^{a \dot{a}}$, $x_{\dot{a} a}$, $\lambda_3^{Aa}$ and $\tilde{\lambda}_{3 B \dot{b}}$ is as follows:
\begin{align}
\mathcal{M}_3^{a \dot{a}} = g_0 m x^{a \dot{a}} + g_1 \lambda_3^{A a} \tilde{\lambda}_{3B}^{\dot{a}} M_A^B + g_2 x^{a \dot{b}} \tilde{\lambda}_{3 A \dot{b}} \tilde{\lambda}_{3 B}^{ \dot{a}} M^{AB} + g_3 x^{b \dot{a}} \lambda_{3 b}^A \lambda_3^{B a} M_{AB} \nonumber \\
+   g_4 x^{a \dot{b}} x_{\dot{b} c} \lambda_3^{Ac} \tilde{\lambda}_{3 B }^{\dot{a}} M^B_A + g_5 x^{a \dot{b}} x^{b \dot{a}} \tilde{\lambda}_{3 A \dot{b}} \lambda_{3 b}^{B} M_B^A
\end{align}
where we have included an extra factor $m$ in the first term for convenience later on. From the available $SU(4)$ tensors ($\epsilon_{ABCD}, \epsilon^{ABCD}, \frac{1}{m} p_2^{AB}, \frac{1}{m} p_{1AB}, \delta^A_B$), we have $M_A^B = \delta_A^B$, $M^{AB} = \frac{1}{m} p_2^{AB}$ and $M_{AB} = \frac{1}{m} p_{1AB}$. Using these and the fact that $x^{a \dot{a}}x_{\dot{a} b} \sim \delta^a_b$, only the first term survives and hence the amplitude of two scalars equal masses and a massless vector is given by:
\begin{align}
\mathcal{M}_3^{a \dot{a}} = g m x^{a \dot{a}}
\end{align}
This amplitude can be easily obtained from the scalar QED interaction: $g\phi A^{\mu}\partial_{\mu}\phi$.

For the next example, let the particle $1$ be massive spin-half particle while particles $2$ and $3$ are massive scalar and massless vector respectively. Then the most general polynomial constructed out of $\lambda_1^{A I}$, $x^{a \dot{a}}$, $x_{\dot{a} a}$, $\lambda_3^{Aa}$ and $\tilde{\lambda}_{3 B \dot{b}}$ is as follows:
\begin{align}
\mathcal{M}_3^{I a \dot{a}} = g_0 \sqrt{m} x^{a \dot{a}} \lambda_1^{A I} M_A + \frac{1}{\sqrt{m}}
( g_1 \lambda_3^{A a} \tilde{\lambda}_{3B}^{\dot{a}} \lambda_1^{C I} M_{AC}^B + g_2 x^{a \dot{b}} \tilde{\lambda}_{3 A \dot{b}} \tilde{\lambda}_{3 B}^{ \dot{a}} \lambda_1^{C I} M^{AB}_C \nonumber \\
 + g_3 x^{b \dot{a}} \lambda_{3 b}^A \lambda_3^{B a} \lambda_1^{C I} M_{AB C}
+   g_4 x^{a \dot{b}} x_{\dot{b} c} \lambda_3^{Ac} \tilde{\lambda}_{3 B }^{\dot{a}} \lambda_1^{C I} M^B_{AC} + g_5 x^{a \dot{b}} x^{b \dot{a}} \tilde{\lambda}_{3 A \dot{b}} \lambda_{3 b}^{B} \lambda_1^{C I} M_{BC}^A )
\end{align}
From the available tensors of $SU(4)$, we can see that none of $M_A, m^{AB}_C, M_{ABC}, M_{AB}^C$ and $M^A_{BC}$ can be constructed. Hence the amplitude corresponding to this process vanishes.

As a third example, let's consider the process involving two massive spin half particles (particle $1$ and $2$) with equal mass $m$ and a massless vector particle containing one dotted and one undotted index. Then the scattering amplitude constructed using $\lambda_1^{AI}, \lambda_2^{BJ}, \lambda_3^{Ca}, \tilde{\lambda}_{3D}^{\dot{a}}$ and $x^{a \dot{a}}$ is given by:
\begin{align}
\mathcal{M}_3^{IJa \dot{a}} = \lambda_1^{AI} \lambda_2^{BJ} \{ g_0m x^{a \dot{a}} M_{AB} + g_1  \lambda_3^{Ca} \tilde{\lambda}_{3D}^{\dot{a}} M_{ABC}^D + g_2  x^{a \dot{b}} \tilde{\lambda}_{3 C \dot{b}} \tilde{\lambda}_{3D}^{\dot{a}} M_{AB}^{CD} \nonumber \\
+ g_3  x^{b \dot{a}} \lambda_{3b}^C \lambda_3^{Da} M_{ABCD}  + g_4  x^{a \dot{b}} x^{b \dot{a}} \tilde{\lambda}_{3C \dot{b}} \lambda_{3b}^D M_{ABD}^C \}
\end{align}
Now by constructing the various $M$ above using the available $SU(4)$ tensors, we get the following result:
\begin{align}
\label{QED}
\mathcal{M}_3^{IJ a \dot{a}} = &g_0 m^2 x^{a \dot{a}} \left( \langle 1^I 2^J] + [1^I 2^J \rangle \right) \nonumber \\
+ &g_1 \left[ \left( \langle 2^J 3^{\dot{a}}] [1^I 3^a \rangle + \langle 1^I 3^{\dot{a}}] [2^J 3^a \rangle \right) + x^{a \dot{b}} \left( \langle 1^I 3^{\dot{a}}] [3_{\dot{b}} 2^J \rangle + 
 \langle 1^I 3_{\dot{b}}][3^{\dot{a}} 2^J \rangle \right) \right. \nonumber \\
 &\hspace{30mm} + \left. y^{\dot{a} b} \left( \langle 3^a 1^I] [2^J 3_b \rangle + \langle 3_b 1^I][2^J 3^a \rangle \right) \right] \nonumber \\
+ &g_2 x^{a \dot{b}}[1^I 2^J 3_{\dot{b}} 3^{\dot{a}}] + g_3 x^{ b \dot{a}} \langle 1^I 2^J 3^a 3_b \rangle + g_4 x^{a \dot{b}} x^{b \dot{a}} \left(  \langle 2^J 3_{\dot{b}}] [1^I 3_b \rangle + \langle 1^I 3_{\dot{b}}] [2^J 3_b \rangle  \right)
\end{align} 
where $g_i$ are  coupling constants that depend on the theory.

We now show that part of the above structures can be obtained from the following minimal coupling terms:
\begin{align}
\bar{\psi}\slashed{A}\psi ={\lambda^{\dagger}}\slashed{A}\lambda +\tilde{\lambda }^{\dagger}\slashed{A}\tilde{\lambda }
\end{align}
Feynman rules give us the following expression:
\begin{align}
\lambda_{1 }^{AI}\lambda_{2 }^{BJ}\varepsilon_{3AB}^{a\dot{a}}+ \tilde{\lambda }_{1A}^I\tilde{\lambda }_{2B}^J\varepsilon^{ABa\dot{a}}_3
\end{align}
To start with, let us take the first term and write it as follows:
\begin{align}
\lambda_{1 }^{AI}\lambda_{2 }^{BJ}\varepsilon_{3AB}^{a\dot{a}}&=\frac{1}{m^2}\left[p_1^{AC}p_2^{BD}-p_1^{BC}p_2^{AD}\right]\tilde{\lambda }_{1C}^I\tilde{\lambda }_{2D}^J\varepsilon_{3AB}^{a\dot{a}}
\end{align}
Now replacing $p_1$ as $-p_2-p_3$ and using the following identity\footnote{This can be derived by first noting that the LHS of the identity is in completely anti-symmetric representation of $SU(4)$ and hence should be proportional to $\epsilon ^{ABCD}$. Proportionality constant can be fixed by contracting with say $p_{AB}$ and using the relation $p^{CD}=\frac{1}{2}\epsilon^{ABCD}p_{AB}$.}:
\begin{align}
-p^{AC}p^{BD}+p^{BC}p^{AD}+p^{AB}p^{CD}=-m^2\epsilon ^{ABCD}
\end{align}
we find that\footnote{In writing the below result, we use the fact that $p_3^{AB}\varepsilon_{3BC}^{A\dot{a}}\sim \lambda _3^{Aa}\tilde{\lambda }_{3C}^{\dot{a}}$. This can be derived just by writing momentum and polarization vector in terms of spinor helicity variables.}
\begin{align}
\lambda_{1 }^{AI}\lambda_{2 }^{BJ}\varepsilon_{3AB}^{a\dot{a}}+ \tilde{\lambda }_{1A}^I\tilde{\lambda }_{2B}^J\varepsilon^{ABa\dot{a}}_3&=\langle 1^I2^J]x^{a\dot{a}}+\langle 1^I3^{\dot{a}}]\langle 3^a2^J]
\end{align}
The above expression can be written more symmetrically as:
\begin{align}
\lambda_{1 }^{AI}\lambda_{2 }^{BJ}\varepsilon_{3AB}^{a\dot{a}}+ \tilde{\lambda }_{1A}^I\tilde{\lambda }_{2B}^J\varepsilon^{ABa\dot{a}}_3&=\left(\langle 1^I2^J]+\langle 2^J1^I]\right)x^{a\dot{a}}+\left(\langle 1^I3^{\dot{a}}]\langle 3^a2^J]+\langle 2^J3^{\dot{a}}]\langle 3^a1^I]\right)
\end{align}
Clearly, this three point function is a special case of \eqref{QED}.

Lastly, we consider the Yukawa amplitude i.e., interaction between two massive fermions with mass $m_f$ and a massless scalar. This amplitude can be written as follows:
\begin{align}
{\cal M}_3^{IJ} &=g\lambda_{1 }^{AI}\lambda_{2 }^{BJ}M_{AB}
\end{align}
Noting that the only possibilities of $M_{AB}$ are ${p_i}_{AB}$, we get:
\begin{align}
{\cal M}_3^{IJ} &= g\langle 1^I2^J]+g'[1^I2^J\rangle
\end{align}
Note that this amplitude does not include any $x$ factors. Also, this amplitude (with $g=g'$) can be reproduced by the usual Yukawa term: $g\bar{\psi}\psi \phi$.

\subsection{$3$-Massive Particles}

Now, let us consider three massive particles that transform in the representations $R_1$, $R_2$ and $R_3$ of $Sp(4)$ respectively and we denote them as $1^{\{I_1\ldots I_{N_1}\}_{R_1}}$, $2^{\{J_1\ldots J_{N_2}\}_{R_2}}$ and $3^{\{K_1\ldots K_{N_3}\}_{R_3}}$. Then the scattering amplitude can be written as follows:
\begin{align}
\label{all3massive}
{\cal M}_3^{\{I_1\ldots I_{N_1}\}_{R_1}\{J_1\ldots I_{N_2}\}_{R_2}\{K_1\ldots K_{N_3}\}_{R_3}}&=\sum _{\alpha}g_{\alpha}~\tilde{\lambda}_{1A_1}^{  I_1} \ldots \tilde{\lambda}_{1 A_{N_1}}^{I_{N_1}}~\tilde{\lambda}_{2B_1}^{  J_1} \ldots \tilde{\lambda}_{2 B_{N_2}}^{J_{N_2}}~\tilde{\lambda}_{3C_1}^{  K_1} \ldots \tilde{\lambda}_{3 C_{N_3}}^{K_{N_3}}\nonumber \\
&\hspace{20 mm}\times M^{\{A_1\ldots A_{N_1}\}_{R_1}\{B_1\ldots B_{N_2}\}_{R_2}\{C_1\ldots C_{N_3}\}_{R_3}}
\end{align}
As before, sum over $\alpha $ is to indicate that the coupling constant can be different for different terms in $M$. The available $SU(4)$ tensors to construct $M$ are as follows:
\begin{align}
\label{tensorsin3massive}
\delta ^A_B; ~\epsilon_{ABCD}; ~\epsilon^{ABCD}; ~\frac{1}{f_1(m_i)}p_1^{AB}; ~\frac{1}{f_2(m_i)}p_2^{AB}; ~\frac{1}{f_3(m_i)}p_3^{AB}
\end{align}
where $f_i$'s are functions of mass dimension $1$. Even though it is a straightforward exercise, the construction of $M$ can get quite tedious as the number of indices increase. 

Let us now give an example where all the three particles transform as a symmetric 2-tensor of $Sp(4)$ i.e., the particles we consider are $1^{(I_1I_2)}$, $2^{(J_1J_2)}$ and $3^{(K_1K_2)}$ where the parentheses denote symmetrization. Also, for simplicity, let us take the masses of all the particles to be same i.e., $m_1=m_2=m_3=m$. The scattering amplitude can be written as follows:
\begin{align}
\label{3-particles-same mass}
{\cal M}_3^{(I_1I_2)(J_1J_2)(K_1K_2)}=\sum _{\alpha}g_{\alpha}
 ~\tilde{\lambda}_{1A_1 }^{I_1} \tilde{\lambda}_{1A_2 }^{I_2}  \tilde{\lambda}_{2B_1 }^{J_1} \tilde{\lambda}_{2B_2 }^{J_2} \tilde{\lambda}_{3C_1 }^{K_1}  \tilde{\lambda}_{3 C_2}^{K_2} ~M^{(A_1A_2)(B_1B_2)(C_1C_2)}
\end{align}
Some of the possible terms in $M$ are as follows:
\begin{align}
1&. ~\frac{1}{m^3}p_1^{A_1B_1}p_2^{A_2C_1}p_3^{B_2C_2}+\left(\text{symmetrizing in}~ \{A_1,A_2\}, \{B_1,B_2\} ~\text{and} ~\{C_1,C_2\}\right)\\
2&. ~\frac{1}{m^3}p_1^{A_1C_1}p_2^{A_2B_1}p_3^{B_2C_2}+\left(\text{symmetrizing in}~ \{A_1,A_2\}, \{B_1,B_2\} ~\text{and} ~\{C_1,C_2\}\right)\\
3&. ~\frac{1}{m^3}p_1^{B_1C_1}p_2^{A_1C_2}p_3^{A_2B_2}+\left(\text{symmetrizing in}~ \{A_1,A_2\}, \{B_1,B_2\} ~\text{and} ~\{C_1,C_2\}\right)
\end{align}

\section{Four Point Scattering Amplitudes}

In order to construct four point tree amplitudes, we start by noting the fact that the four point function has at most simple poles in the Mandelstam variables $s,t,u$. We construct four point function such that its residues $R_s,R_t,R_u$ in $s,t,u$ channels   is equal to  product of the left and right three point functions. This way of constructing four point functions starting from three point functions can not fix the contact terms. For instance, in the four point amplitude \eqref{4scalars} of 4 scalars, one can in principle add a contact term of the form $\lambda \phi ^4$. Since the three point functions do not know about this contact term, they can not reproduce this. Note that this limitation extends to BCFW recursion relations too as their input is again three point functions.

Before we start, we set the notation for various channels. Define $s_{ij} \coloneqq (p_i + p_j)^2$ which in $4-$particles processes yield the following:
\begin{equation}
\label{channels}
 s=s_{12}, t=s_{24}, u=s_{23}
\end{equation}

The first example we consider is the Yukawa interaction. More precisely, we consider the amplitude of two massless scalars and two massive fermions with mass $m_f$. The possible channels are $s$ and $u$ and are given as follows:
\begin{figure}[htbp!]
	\centering
	\includegraphics[trim={0.2cm 0.2cm 0.2cm 0.5cm},clip,scale=0.4]{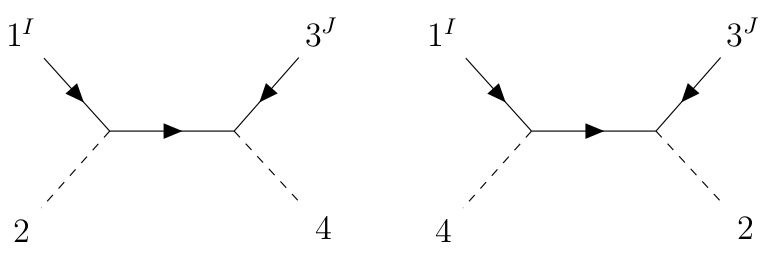}
\end{figure}\\
The residue in $s$-channel is given by the product of the following three point functions:
\begin{figure}[htbp!]
	\centering
	\includegraphics[trim={0.2cm 0.2cm 0.2cm 0.5cm},clip,scale=0.4]{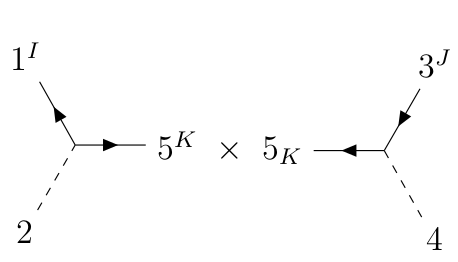}
\end{figure}\\
Using the result of the Yukawa three point amplitudes, we get the residue as
\begin{align}
{\cal R}_s&= \left(g\langle 1^I5^K]+g'[1^I5^K\rangle\right) \left(g\langle 3^J5_K]+g'[3^J5_K\rangle\right)
\end{align}
Using $p_1+p_2=p_5$ and simplifying, we get:
\begin{align}
{\cal R}_s&=m_f[1^I3^J\rangle (g^2+gg')+m_f[3^J1^I\rangle (g'^2+gg')+g^2\langle 1^I|2|3^J\rangle +g'^2[1^I|2|3^J]
\end{align}
Similarly, the residue in the $u$-channel is given by the following:
\begin{figure}[htbp!]
	\centering
	\includegraphics[trim={0.2cm 0.2cm 0.2cm 0.5cm},clip,scale=0.4]{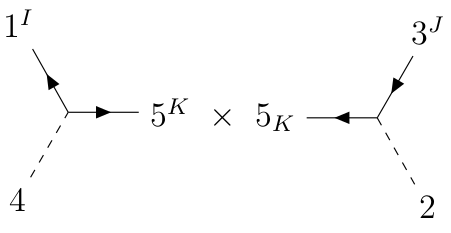}
\end{figure}\\


Using $p_1+p_4=p_5$, we get:
\begin{align}
{\cal R}_u&=m_f[1^I3^J\rangle (g^2+gg')+m_f[3^J1^I\rangle (g'^2+gg')+g^2\langle 1^I|4|3^J\rangle +g'^2[1^I|4|3^J]
\end{align}
Since there are no spurious poles in these residues, the four amplitude is simply given as:
\begin{align}
{\cal M}_4^{IJ}=\frac{{\cal R}_s}{s-m_f^2}+\frac{{\cal R}_u}{u-m_f^2}
\end{align}

Next example is $f\bar{f}\rightarrow \bar{f}f$ which again concerns with Yukawa interaction. In the $s$-channel, Feynman diagram is given as follows:
\begin{figure}[htbp!]
	\centering
	\includegraphics[trim={0.2cm 0.15cm 0.2cm 0.5cm},clip,scale=0.4]{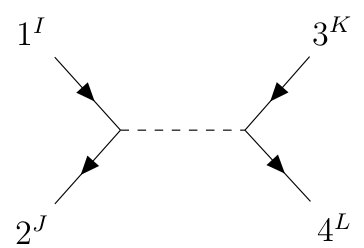}
\end{figure}\\
Scalar propagator is massless here and particles $1,2,3,4$ are massive with all having equal mass $m$. Then the residue in the $s-$channel is given as follows:
\begin{equation}
{\cal R}_s = g^2 \left( [1^I 2^J \rangle + \langle 1^I 2^J] \right) \left( \langle 4^L 3^K] + [4^L  3^K \rangle \right)
\end{equation}
Similarly in the $t-$channel, we have the following Feynman diagram:
\begin{figure}[htbp!]
	\centering
	\includegraphics[trim={0.2cm 0.15cm 0.2cm 0.3cm},clip,scale=0.35]{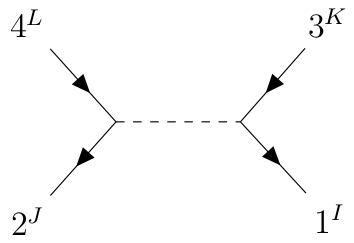}
\end{figure}\\
\newpage
Residue in this channel is given as follows:
\begin{equation}
{\cal R}_t = g^2 \left( [4^L 2^J \rangle + \langle 4^L 2^J] \right) \left( \langle 1^I 3^K] + [1^I  3^K \rangle \right)
\end{equation}
Since there are no spurious poles appearing, the four point scattering amplitude is given as follows:
\begin{align}
{\cal M}_4^{IJKL}=\frac{{\cal R}_s}{s}+\frac{{\cal R}_t}{t}
\end{align}

As a third example, we consider analogue of Compton scattering for scalars in $6D$ i.e., we consider $\phi \gamma \rightarrow \phi \gamma $. This process includes $s$ and $u$ channels and are given as:
\begin{figure}[htbp!]
	\centering
	\includegraphics[trim={0.2cm 0.2cm 0.2cm 0.5cm},clip,scale=0.4]{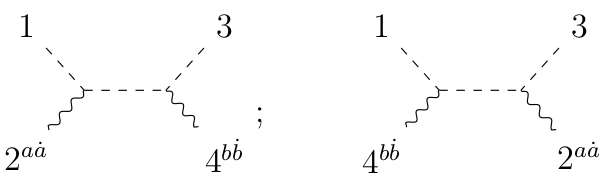}
\end{figure}\\
Here particles $1$ and $3$ are massive scalars along with a massive scalar propagator (all of equal mass $m$) and particles $2$ and $4$ are massless vectors. Then the product of the residues in the above $s-$channel is as follows:
\begin{equation}
\mathcal{R}_s =  g^2m^2 x_{12}^{a \dot{a}}x_{34}^{b \dot{b}}
\end{equation}
To simplify this expression, we start by writing down the following equations:
\begin{align}
p_{IAB}\lambda_{2 }^{Aa}&=mx^{a\dot{a}}_{12}\tilde{\lambda }_{2B\dot{a}}; ~~ p_I^{CB}\lambda _{4C}^{\dot{b}}=-mx_{34}^{\dot{b}b}\lambda _{4b}^{B}
\end{align}
where $p_I$ denotes the momentum of the scalar propagator. Multiplying these two equations, we get:
\begin{align}
\langle 2^a4^{\dot{b}}]&=x_{12}^{a \dot{a}}x_{34}^{b \dot{b}} ~\langle 4_b2_{\dot{a}}]
\end{align}
Using the following identity:
\begin{align}
\langle 2^a4^{\dot{b}}] \langle 2^c4^{\dot{d}}]^{-1} -\langle 2_c4^{\dot{b}}] \langle 2_a4^{\dot{d}}]^{-1}&=-\delta _c^a \delta ^{\dot{b}}_{\dot{d}}
\end{align}
we find the residue in $s$-channel to be:
\begin{align}
{\cal R}_s&=g^2\frac{m^2\langle 4^b2^{\dot{a}}]\langle 2^a4^{\dot{b}}]-\langle 4^b|1|2^a\rangle  [4^{\dot{b}}|3|2^{\dot{a}}]}{t}
\end{align}
Note that the $s$-channel residue contains a pole in $t$. Similarly, we can compute the residue along $u$-channel and the four point amplitude that reproduces these residues is:
\begin{align}
\label{scalar compton}
{\cal M}^{a\dot{a};b\dot{b}}&=g^2\frac{m^2\langle 4^b2^{\dot{a}}]\langle 2^a4^{\dot{b}}]-\langle 4^b|1|2^a\rangle  [4^{\dot{b}}|3|2^{\dot{a}}]}{(s-m^2)(u-m^2)}
\end{align} 
While computing residues, we need to use the relation $s+t+u=2m^2$. For instance, we can see that in the $s$-channel, the pole is at $s=m^2$ and near this pole, we have $m^2 - u = t$. Using this, we recover the correct residue.

As a fourth example, we consider the interaction $\phi \phi ^* \rightarrow \phi \phi ^*$ which is mediated by a photon. This process has $s$ and $u$ channels and are given as following:
\begin{figure}[htbp!]
	\centering
	\includegraphics[trim={0.2cm 0.2cm 11.7cm 0.3cm},clip,scale=0.4]{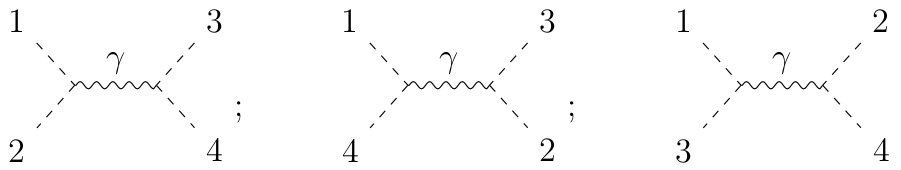}
\end{figure}\\
Here particles $1, 2, 3, 4$ are scalars with each of them having mass $m$ while the propagator is massless vector. We start with the residue in the $s$-channel that is given by: 
\begin{equation}
\mathcal{R}_s = g^2 m^2   x_{12}^{a \dot{a}}   {x_{34}}_{\dot{a} a}=-g^2\frac{\langle \zeta ^c|2|5^a \rangle}{\langle \zeta ^c5_{\dot{a}}]}\frac{[\chi ^{\dot{c}}|3|5_{\dot{a}}]}{[\chi  ^{\dot{c}} 5^a\rangle }
\end{equation}
where $\zeta $ and $\chi $ are reference spinors and $g$ is a dimensionless constant. To simplify this expression, we need the following identity:
\begin{align}
p_{AB}q^{CD}&=-2(p.q)\left(\delta_A ^C\delta_B^D-\delta_A ^D\delta_B^C\right)+p^{CD}q_{AB}+\left(\delta_A ^Cq_{BF}p^{FD}-\delta_A ^Dq_{BF}p^{FC}-(A\leftrightarrow B)\right) 
\end{align}
Using this identity along\footnote{Note that the following identity can be obtained by multiplying the following two equations:
\begin{align}
p_{2AB}\lambda _5^{Aa}&=mx_{12}^{a\dot{a}}\tilde{\lambda }_{5B\dot{a}}; ~~ p_3^{CB}\tilde{\lambda } _{5C}^{\dot{a}}=my_{34\dot{a}b}\lambda _5^{Ba}
\end{align}} with $\langle 5^a|2.3|5^{\dot{a}}]=0$, we obtain the residue as:
\begin{align}
{\cal R}_s &=2g^2 (p_2.p_3)
\end{align}
Note that contrary to previous examples, there are no poles in $t$ or $u$ even though there are $x$-factors. Further, if we have started the computation of ${\cal R}_s$ by writing $x_{12}^{a\dot{a}}=-\frac{\langle \zeta ^c|1|5^a \rangle}{\langle \zeta ^c5_{\dot{a}}]}$, then we get ${\cal R}_s=-2g^2(p_1.p_3)$. So, more symmetrically, we write the residue as follows:
\begin{align}
{\cal R}_s&=g^2(p_2-p_1).p_3
\end{align} 

The residue in the $u$-channel is obtained by replacing $1\leftrightarrow 3$ in ${\cal R}_s$. Thus, the four point amplitude of this process is given by:
\begin{align}
\label{4scalars}
{\cal M}_4&=g^2\left(\frac{(p_2-p_1).p_3}{s}+\frac{(p_2-p_3).p_1}{u}\right)
\end{align} 


\section{Higgs Mechanism}

In this section, we discuss how Higgs mechanism can be understood by demanding that the massive amplitudes reduces to massless amplitudes in the high energy limit. We take the three point amplitude of massive vectors as working example. In the high energy limit, we demand that it should reduce to sum of amplitudes of the following diagrams:
	\begin{figure}[htbp!]
	\centering
	\includegraphics[trim={0.2cm 0.2cm 0.2cm 0.3cm},clip,scale=0.4]{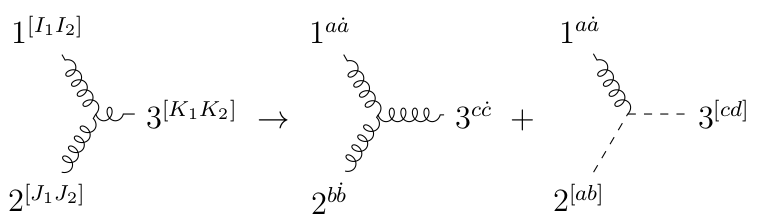}
	\caption{High energy limit of $3-$point massive vector amplitude}
	\label{massive to massless}
\end{figure}
\\
Note that since $a,b,\ldots $ are $SU(2)$ indices, $2^{[ab]}$ and $3^{[cd]}$ are scalar particles with one independent component each. 

Our strategy would be to retain those terms in massive 3-vector amplitude which would exactly reproduce the massless amplitudes of both the diagrams on RHS of figure \ref{massive to massless} in the high energy limit. After retaining such terms in massive amplitude, we consider the relevant components and explicitly show that in the high energy limit, we reproduce the 3-gluon interaction and the vector-scalar-scalar interaction.

We start by constructing the amplitude of three massive vectors that are labelled\footnote{We are taking the antisymmetric representation of the little group as it has five independent components that match with the degrees of freedom  of a massive vector in $6D$.} as $1^{[I_1 I_2]}$, $2^{[J_1 J_2]}$ and $3^{[K_1 K_2]}$ using the strategy presented in the section \ref{3-points}. The massive scattering amplitude of $1^{[I_1 I_2]}$, $2^{[J_1 J_2]}$ and $3^{[K_1 K_2]}$ is given as follows:
\begin{align}
\label{3-particles-same mass}
{\cal M}_3^{[I_1 I_2]; [J_1 J_2]; [K_1 K_2]}=\sum _{\alpha}\frac{g_{\alpha}}{m^2}
\lambda_{1 }^{A_1 I_1} \lambda_{1 }^{A_2 I_2}  \lambda_{2 }^{B_1 J_1} \lambda_{2}^{B_2 J_2} \lambda_{3 }^{C_1 K_1}  \lambda_{3 }^{C_2 K_2} ~M_{[A_1 A_2]; [B_1 B_2]; [C_1 C_2]}
\end{align}
where a factor of $m^2$ is used to make the coupling constants $g_{\alpha}$ dimensionless\footnote{Note that we have used the fact that the dimension of three vector amplitude is one. See \cite{Cheung} for example.}. Using various tensor structures, we obtain the above amplitude as a function of following Lorentz ($SU(4)$) invariant and the little group ($Sp(4)$) covariant quantities: 
\begin{align}
&\langle 1^{I_1}1^{I_2}]; ~\langle 2^{J_1}2^{J_2}];  ~\langle 3^{K_1}3^{K_2}] \nonumber \\
&\langle 1^{I_1}2^{J_2}];  ~\langle 2^{J_1}3^{K_2}];  ~\langle 3^{K_1}1^{I_2}]; ~\langle 1^{I_1}3^{K_2}]; ~\langle 2^{J_1}1^{I_2}]; ~\langle 3^{K_1}2^{J_2}] \nonumber \\
&\langle 1^{I_2}2^{J_1}];  ~\langle 2^{J_2}3^{K_1}];  ~\langle 3^{K_2}1^{I_1}]; ~\langle 1^{I_2}3^{K_1}]; ~\langle 2^{J_2}1^{I_1}]; ~\langle 3^{K_2}2^{J_1}]\nonumber \\
&\langle 1^{I_1}2^{J_1}];  ~\langle 2^{J_1}3^{K_1}];  ~\langle 3^{K_1}1^{I_1}]; ~\langle 1^{I_1}3^{K_1}]; ~\langle 2^{J_1}1^{I_1}]; ~\langle 3^{K_1}2^{J_1}]\nonumber \\
&\langle 1^{I_2}2^{J_2}];  ~\langle 2^{J_2}3^{K_2}];  ~\langle 3^{K_2}1^{I_2}]; ~\langle 1^{I_2}3^{K_2}]; ~\langle 2^{J_2}1^{I_2}]; ~\langle 3^{K_2}2^{J_2}]\nonumber \\
&\langle 1^{I_1}|2|1^{I_2}\rangle ; ~\langle 2^{J_1}|3|2^{J_2}\rangle; ~\langle 3^{K_1}|1|3^{K_2}\rangle ; ~ [1^{I_1}|2|1^{I_2}]; ~ [2^{J_1}|3|2^{J_2}]; ~[3^{K_1}|1|3^{K_2}]\nonumber \\
&\langle 1^{I_1}1^{I_2}2^{J_1}2^{J_2}\rangle ;~\langle 1^{I_1}1^{I_2}3^{K_1}3^{K_2}\rangle ;~\langle 2^{J_1}2^{J_2}3^{K_1}3^{K_2}\rangle ; ~[1^{I_1}1^{I_2}2^{J_1}2^{J_2}];~[1^{I_1}1^{I_2}3^{K_1}3^{K_2}]; ~[2^{J_1}2^{J_2}3^{K_1}3^{K_2}]\nonumber \\
&\langle 1^{I_1}1^{I_2}2^{J_1}3^{K_1}\rangle ;~\langle 1^{I_1}2^{J_1}2^{J_2}3^{K_1}\rangle ;~\langle 1^{I_1}2^{J_1}3^{K_1}3^{K_2}\rangle ;~[1^{I_1}1^{I_2}2^{J_1}3^{K_1}]; ~[1^{I_1}2^{J_1}2^{J_2}3^{K_1}]; ~[1^{I_1}2^{J_1}3^{K_1}3^{K_2}]\nonumber \\
&\langle 1^{I_1}1^{I_2}2^{J_2}3^{K_1}\rangle ;~\langle 1^{I_2}2^{J_1}2^{J_2}3^{K_1}\rangle ;~\langle 1^{I_2}2^{J_1}3^{K_1}3^{K_2}\rangle ;~[1^{I_1}1^{I_2}2^{J_2}3^{K_1}]; ~[1^{I_2}2^{J_1}2^{J_2}3^{K_1}]; ~[1^{I_2}2^{J_1}3^{K_1}3^{K_2}]\nonumber \\
&\langle 1^{I_1}1^{I_2}2^{J_1}3^{K_2}\rangle ;~\langle 1^{I_1}2^{J_1}2^{J_2}3^{K_2}\rangle ;~\langle 1^{I_1}2^{J_2}3^{K_1}3^{K_2}\rangle ;~[1^{I_1}1^{I_2}2^{J_1}3^{K_2}]; ~[1^{I_1}2^{J_1}2^{J_2}3^{K_2}]; ~[1^{I_1}2^{J_2}3^{K_1}3^{K_2}]\nonumber \\
&\langle 1^{I_1}1^{I_2}2^{J_2}3^{K_2}\rangle ;~\langle 1^{I_2}2^{J_1}2^{J_2}3^{K_2}\rangle ;~\langle 1^{I_2}2^{J_2}3^{K_1}3^{K_2}\rangle ;~[1^{I_1}1^{I_2}2^{J_2}3^{K_2}]; ~[1^{I_2}2^{J_1}2^{J_2}3^{K_2}]; ~[1^{I_2}2^{J_2}3^{K_1}3^{K_2}]
\end{align}
where we used the following notation:
\begin{align}
\langle i^Ij^J]&=\lambda_{i}^{AI}\tilde{\lambda } _{jA}^J; ~\langle i^I|p|j^J\rangle =\lambda_{i}^{AI}\lambda  _{j}^{BJ}p_{AB}; ~\langle i^Ij^Jk^Kl^L\rangle =\epsilon_{ABCD}\lambda_{i}^{AI}\lambda_{j}^{BJ}\lambda_{k}^{CK}\lambda_{l}^{DL} 
\end{align}
The other possible structures that we did not mention here can be written down using the listed quantities.

Now, we demand that the above three point amplitude of massive vectors matches with the appropriate amplitudes in the high energy limit. As a result of this constraint, we need to keep only the following terms in the massive amplitude:
\begin{align}
\label{3-massive}
{\cal M}&^{[I_1I_2];[J_1J_2];[K_1K_2]}_3\nonumber \\
&=\frac{1}{m^3}\left(\langle 3^{[K_1}|p_1-p_2|3^ {K_2]}\rangle \right)\left(\langle 1^{I_1}2^{J_1}]\langle 1^ {I_2}2^{J_2}]-\langle 1^{I_1}2^{J_2}]\langle 1^ {I_2}2^{J_1}]\right)\nonumber \\
&+\frac{1}{m^3}\left(\langle 2^{[J_1}|p_3-p_1|2^ {J_2]}\rangle \right) \left(\langle 1^{I_1}3^{K_1}]\langle 1^ {I_2}3^{K_2}]-\langle 1^{I_1}3^{K_2}]\langle 1^ {I_2}3^{K_1}]\right)\nonumber \\
&+\frac{1}{m^3}\left(\langle 1^{[I_1}|p_2-p_3|1^ {I_2]}\rangle \right) \left(\langle 2^{J_1}3^{K_1}]\langle 2^ {J_2}3^{K_2}]-\langle 2^{J_1}3^{K_2}]\langle 2^ {J_2}3^{K_1}]\right)\nonumber \\
&-\text{terms containing}~J^{I_1I_2}~\text{and/or}~J^{J_1J_2}~\text{and/or}~J^{K_1K_2}
\end{align}
The last line is needed to remove the trace terms from the amplitude. We did not explicitly write them down as they vanish while choosing the component $e^{+I_1}_ae^{-I_2}_{\dot{a}}e^{+J_1}_be^{-J_2}_{\dot{b}}e^{+K_1}_ce^{-K_2}_{\dot{c}}$. As will be explained later, this component is relevant while taking the high-energy limit to match it with the 3-gluon amplitude of \cite{Cheung}. These trace components are important while considering the vector-scalar-scalar interaction and we write those terms explicitly when we need them. 

This amplitude can also be obtained from the interaction of the form $(\partial _{\mu} A_{\nu}) A^{\mu}A^{\nu}$ using Feynman rules as follows: 
\begin{align}
V^{[I_1I_2];[J_1J_2];[K_1K_2]}&=\left(\varepsilon _1^{[I_1I_2]}.(p_2-p_3)\right) (\varepsilon _2^{[J_1J_2]}.\varepsilon _3^{[K_1K_2]})+\text{cyclic in 1,2,3 particles}
\end{align}
where the massive polarization vectors are given as follows:
\begin{align}
\label{massive polarization}
(\varepsilon _i^{[IJ]})^{AB}&=\frac{1}{m}\left(\lambda_{i} ^{AI} \lambda_{i}^{BJ}-\lambda_{i} ^{BI} \lambda_{i}^{AJ}+ \frac{1}{2}p_i^{AB}J^{IJ}\right)\nonumber \\
(\varepsilon _i^{[IJ]})_{AB}&=\frac{1}{m}\left(\tilde{\lambda}_{iA} ^{I} \tilde{\lambda}_{iB}^{J}-\tilde{\lambda}_{iB} ^{I} \tilde{\lambda}_{iA}^{J}- \frac{1}{2}p_{iAB}J^{IJ}\right)\nonumber \\
(\varepsilon _{[IJ]})^{AB}&=\frac{1}{m}\left(\lambda_{I}^A \lambda_{J}^B-\lambda_{I}^B \lambda_{J}^A- \frac{1}{2}p^{AB}J_{IJ}\right)\nonumber \\
(\varepsilon _{[IJ]})_{AB}&=\frac{1}{m}\left(\tilde{\lambda}_{iAI} \tilde{\lambda}_{iBJ}-\tilde{\lambda}_{iBI} \tilde{\lambda}_{iAJ}+ \frac{1}{2}p_{iAB}J_{IJ}\right)
\end{align}
The second form can be obtained\footnote{The following identity  is useful in deriving this:
\begin{align}
\epsilon_{ABCD} \lambda ^{CI}\lambda^{DJ}&=\tilde{\lambda} _A ^{I}\tilde{\lambda} _B^{J}-\tilde{\lambda} _A^{J}\tilde{\lambda} _B ^{I}-p_{AB}J^{IJ}
\end{align}} by using the relation $\varepsilon _{AB}=\frac{1}{2}\epsilon_{ABCD}\varepsilon^{CD}$ and third one by contracting the first equation with $J_{KI}J_{LJ}$. Note that this choice of polarization vectors satisfy the following necessary relations:
\begin{align}
\varepsilon_{iAB} p^{AB}_i&=\varepsilon^{AB}_i p_{iAB}=0 \nonumber \\
(\varepsilon _i^{[IJ]})^{AB}(\varepsilon _i^{[KL]})_{AB}&=2\left(J^{IK}J^{JL}-J^{JL}J^{IK}\right)-J^{IJ}J^{KL}\nonumber \\
(\varepsilon _i^{[IJ]})^{AB}(\varepsilon _{i[IJ]})_{CD}&=2\left(\delta ^A_C\delta ^B_D-\delta ^A_D\delta ^B_C\right)+\frac{1}{m^2}p^{AB}p_{CD}
\end{align}

We now show that in the high energy limit this massive amplitude indeed reduces to the appropriate amplitudes as given in the figure \ref{massive to massless}.  First, we consider the high energy limit of the component $e^{+I_1}_ae^{-I_2}_{\dot{a}}e^{+J_1}_be^{-J_2}_{\dot{b}}e^{+K_1}_ce^{-K_2}_{\dot{c}}$. This should reproduce the three point amplitude of gluons considered in \cite{Cheung}. The corresponding amplitude of the above component is as follows:
\begin{align}
\label{terms retained in HE limit}
{\cal M}^{a\dot{a};b\dot{b};c\dot{c}}=\frac{1}{p_1p_2p_3}&\left[ \langle 3^c|1|\eta_{3}^ {\dot{c}}\rangle \left(\langle 1^a\tilde{\eta} _2^{b}]\langle \eta_{1}^ {\dot{a}}2^{\dot{b}}]-\langle 1^a2^{\dot{b}}]\langle \eta _1^{\dot{a}}\tilde{\eta} _{2}^b]\right)+\text{cyclic in 1,2,3 particles}\right]
\end{align}
where $p_{i}^2=E_i^2-m^2$. More suggestively, we can also write this expression as:
\begin{align}
{\cal M}^{a\dot{a};b\dot{b};c\dot{c}}=\frac{\langle 3^c|1|\eta_{3}^ {\dot{d}}\rangle}{\langle \eta_{3}^ {\dot{d}}3_{\dot{c}}]}\left(\frac{\langle 1^a|\bar{\sigma}^{\mu} |\eta_{1}^ {\dot{d}}\rangle}{\langle \eta_{1}^ {\dot{d}}1_{\dot{a}}]}\frac{[ 2^{\dot{b}}|\sigma_{\mu} |\tilde{\eta}_{2}^ {d}]}{[\tilde{\eta}_{2}^ {d} 2_{b}]}\right)+\text{cyclic in 1,2,3 particles}
\end{align}
This expression is same as the three gluon amplitude in equation (52) of \cite{Cheung} if we choose the reference spinors in the polarization vectors of $i^{\text{th}}$ particle to be $\eta _i$. As the final answer should not depend on the choice of polarization vectors, we see that we have obtained the 3-gluon amplitude by taking the high energy limit  of the massive amplitude \eqref{3-massive}. Using KLT relations \cite{KLT} we accordingly obtain the $3-$point amplitude in linearized gravity.

Note that this can also be obtained using the Feynman rules computation of this particular component coming from the three point vertex similar to gluons. More explicitly, this component of three point vertex can be computed using Feynman rules as follows:
\begin{align}
V^{a\dot{a};b\dot{b};c\dot{c}}&=\left(\varepsilon _1^{a\dot{a}}.(p_2-p_3)\right) (\varepsilon _2^{b\dot{b}}.\varepsilon _3^{c\dot{c}})+\left(\varepsilon _2^{b\dot{b}}.(p_3-p_1)\right) (\varepsilon _3^{c\dot{c}}.\varepsilon _1^{a\dot{a}})+\left(\varepsilon _3^{c\dot{c}}.(p_1-p_2)\right) (\varepsilon _1^{a\dot{a}}.\varepsilon _2^{b\dot{b}})
\end{align}
where the corresponding components of massive polarization vectors are as follows:
\begin{align}
(\varepsilon_i ^{a\dot{a}})^{AB}&=\frac{1}{p_i}\left[\lambda_{i }^{Aa}\eta _i^{B\dot{a}}-\lambda_{i }^{Ba}\eta _i^{A\dot{a}}\right]\nonumber \\
(\varepsilon_i ^{a\dot{a}})_{AB}&=\frac{1}{p_i}\left[\tilde{\eta} _{iB}^{a}\tilde{\lambda}_{iA }^{\dot{a}}-\tilde{\eta} _{iA}^{a}\tilde{\lambda}_{iB }^{\dot{a}}\right]
\end{align}
Using these expressions, we can show that $V^{a\dot{a};b\dot{b};c\dot{c}}$ is same as the terms we retained in \eqref{terms retained in HE limit} up to a numerical factor.

We showed that the 3-gluon amplitude can be reproduced by taking the high energy limit of certain component of the massive amplitude\eqref{3-massive}. Now, we show that the appropriate component of massive amplitude \eqref{3-massive} also reproduces correctly the scalar-scalar-vector amplitude. In this case, we need to choose the  $e^{+I_1}_ae^{-I_2}_{\dot{a}}e^{+J_1}_{[b_1}e^{+J_2}_{b_2]}e^{-K_1}_{[\dot{c}_1}e^{-K_2}_{{\dot{c}}_2]}$ component of the massive amplitude \eqref{3-massive}.

For convenience, let us rewrite \eqref{3-massive} here:
\begin{align}
\label{3-massive repeat}
{\cal M}&^{[I_1I_2];[J_1J_2];[K_1K_2]}_3\nonumber \\
&=\frac{1}{m^3}\left(\langle 3^{[K_1}|p_1-p_2|3^ {K_2]}\rangle \right)\left(\langle 1^{I_1}2^{J_1}]\langle 1^ {I_2}2^{J_2}]-\langle 1^{I_1}2^{J_2}]\langle 1^ {I_2}2^{J_1}]\right)\nonumber \\
&+\frac{1}{m^3}\left(\langle 2^{[J_1}|p_3-p_1|2^ {J_2]}\rangle \right) \left(\langle 1^{I_1}3^{K_1}]\langle 1^ {I_2}3^{K_2}]-\langle 1^{I_1}3^{K_2}]\langle 1^ {I_2}3^{K_1}]\right)\nonumber \\
&+\frac{1}{m^3}\left(\langle 1^{[I_1}|p_2-p_3|1^ {I_2]}\rangle \right) \left(\langle 2^{J_1}3^{K_1}]\langle 2^ {J_2}3^{K_2}]-\langle 2^{J_1}3^{K_2}]\langle 2^ {J_2}3^{K_1}]\right)\nonumber \\
&-\text{terms containing}~J^{I_1I_2}~\text{and/or}~J^{J_1J_2}~\text{and/or}~J^{K_1K_2}
\end{align}
As already mentioned in the beginning of the section, one difference with the previous case is that the relevant components of $J^{J_1J_2}/J^{K_1K_2}$ are not zero and we need to consider those terms. We write those terms explicitly when we need them. 

Let us start by looking at the term\footnote{Note that  $J_{J_1J_2}\langle 2^{[J_1}|p_3-p_1|2^ {J_2]}\rangle =0$ as $2p_2.p_3=2p_1.p_3=-m^2$ and hence there is no $J$ factor in this term.} in first bracket of second line.  Taking the $e^{+J_1}_{[b_1}e^{+J_2}_{b_2]}$ component of $\langle 2^{[J_1}|p_3-p_1|2^ {J_2]}\rangle$ and noting the identity\footnote{Note that this identity is true only for massless particle $i$.} $\langle p_i^a|q|p_i^b\rangle =-2(p_i.q)\epsilon^{{a}{b}}$, we see that the term is zero in the limit $m\rightarrow 0$. Similarly, we can show that the first line also does not contribute.

Now consider the third line. This can be written along with terms with $J$ factors as:
\begin{align}
\frac{1}{m^3}\left(\langle 1^{[I_1}|p_2-p_3|1^ {I_2]}\rangle \right) &\left(\langle 2^{J_1}3^{K_1}]\langle 2^ {J_2}3^{K_2}]-\langle 2^{J_1}3^{K_2}]\langle 2^ {J_2}3^{K_1}]-\frac{1}{2}J^{K_1K_2}\langle 2^{J_1}|3|2^{J_2}\rangle \right.\nonumber \\
&\hspace{38mm}\left. +\frac{1}{2}J^{J_1J_2}[3^{K_1}|2|3^{K_2}]-\frac{m^2}{4}J^{J_1J_2}J^{K_1K_2}\right)
\end{align}
Consider the terms in second bracket. To begin with, we first note the following identity\footnote{It can be easily proved starting from $\langle p_i^{I_1}p_i^{I_2}]=-m_iJ^{I_1I_2}$.}:
\begin{align}
J^{IJ}&=\epsilon^{{a}{b}} e^{+I}_{a}e^{+J}_{b}+\epsilon^{\dot{a} \dot{b}}e^{-I}_{\dot{a}}e^{-J}_{{\dot{b}}}
\end{align}
In the limit $m\rightarrow 0$, using this identity along with $\langle p_i^a|q|p_i^b\rangle =-2(p_i.q)\epsilon^{{a}{b}}$ immediately shows that the $e^{+J_1}_{[b_1}e^{+J_2}_{b_2]}e^{-K_1}_{[\dot{c}_1}e^{-K_2}_{{\dot{c}}_2]}$ component of the last three terms just gives us $ m^2\epsilon^{{b_1}{b_2}}\epsilon^{\dot{c}_1 \dot{c}_2}$ apart from an unimportant numerical factor. 

Regarding the first two terms, the relevant component is:
\begin{align}
\langle 2^{b_1}3^{\dot{c}_1}]\langle 2^{b_2}3^{\dot{c}_2}]-\langle 2^{b_2}3^{\dot{c}_1}]\langle 2^{b_1}3^{\dot{c}_2}]
\end{align}
In the $m\rightarrow 0$ limit, from the momentum conservation we can see that the matrices $\langle 1^{a}2^{\dot{b}}],\langle 2^{b}3^{\dot{c}}],\ldots $ become rank 1 i.e., their determinant becomes zero. Hence, we can write each of these matrices as a product of two $2\times 1 $ matrices. Following the notation of \cite{Cheung}, we write the matrix $\langle 2^{b}3^{\dot{c}}]$ as follows:
\begin{align}
\langle 2^{b}3^{\dot{c}}]&=u_2^b\tilde{u}_3^{\dot{c}}
\end{align} 
Using this expression, it is easy to see that the terms $\langle 2^{b_1}3^{\dot{c}_1}]\langle 2^{b_2}3^{\dot{c}_2}]-(b_1\leftrightarrow b_2)$ becomes zero in $m\rightarrow 0$ limit.

Combining all these, we now see that the $e^{+I_1}_ae^{-I_2}_{\dot{a}}e^{+J_1}_{[b_1}e^{+J_2}_{b_2]}e^{-K_1}_{[\dot{c}_1}e^{-K_2}_{{\dot{c}}_2]}$ component of the massive amplitude \eqref{3-massive repeat} is simply given as\footnote{We are not writing the factors $\epsilon ^{ab}\epsilon^{\dot{a} \dot{b}}$ as they transform in trivial representation of the corresponding little group.}:
\begin{align}
{\cal M}^{a\dot{a}}&=\frac{\langle 1^a|p_2-p_3|\eta_1 ^{\dot{b}}\rangle}{\langle \eta_1 ^{\dot{b}}1_{\dot{a}}]}
\end{align}
This three point function can be reproduced from $\varepsilon _1.(p_2-p_3)$ (i.e., from an interaction of the form $\phi A^{\mu}\partial_{\mu} \phi$) where we choose the reference spinor in the polarization vector to be $\eta _1$. 

To summarize, the terms we retained in the amplitude \eqref{3-massive} of three massive vectors have a good high energy limit. That is, this massive amplitude should reduce to the interactions present in RHS of figure \ref{massive to massless} in the high energy limit. We explicitly showed that this expectation is true by choosing appropriate components of this massive amplitude and reproducing both the 3-gluon interaction (considered in \cite{Cheung}) and vector-scalar-scalar interaction in the high energy limit.

\section{Discussions}

In this paper, we showed that a natural spinor helicity formalism exists for massive particles in six dimensions. We also showed that all the possible structures of three point functions can be completely determined (apart from theory dependent coupling constants) just by demanding the little group covariance and Lorentz invariance. 

Studying amplitudes in six dimensions is interesting in its own right but let us now discuss how these amplitudes play a role elsewhere. For instance in \cite{6d to 4d loops}, the massless spinor helicity formalism in 6d	(along with the associated DHS superspace\cite{DHS}) was useful to determine the dimensionally regularized loop amplitudes in 4d such as one-loop four point function in QCD and multiloop amplitudes in ${\cal N}=4$ SYM. It might be interesting to explore the connection of our massive formalism in 6d to 4d amplitudes.  Further, since the massless amplitudes can be constructed in arbitrary dimensions (see \cite{all dimensions}), it will be interesting to understand the connection between the dimensionally reduced 7d or 8d massless amplitudes and the 6d massive amplitudes we have constructed here in this paper. We leave such explorations to future work.

In the present paper, we constructed various four point functions by demanding that the residues $R_{s,t,u}$ in the corresponding channels are given by product of left and right three point functions. Now, we will conclude by briefly discussing BCFW relations to compute four point functions. BCFW recursion relations\cite{bcfw1} are an immensely useful tool (especially in the case of gauge theories thanks to proliferation of Feynman diagrams) to construct higher ($\geq 4$) point amplitudes starting from three point functions. To find the recursion relations for a $n$-point function denoted by ${\cal A}_n$, we start by choosing any two momenta (say $p_i$ and $p_j$) and shift them as follows:
\begin{equation}
\hat{p}_i^{\mu}(z) = p_i^{\mu} + z q ^{\mu}, \hspace{7mm} \hat{p}_j^{\mu}(z) = p_j^{\mu} - z q^{\mu}
\end{equation}
where $z$ is a complex number and $q^{\mu}$ is independent of $z$. We choose $q^{\mu}$ such that $\hat{p}_i^2=p_i^2$ and  $\hat{p}_j^2=p_j^2$. The $n$-point amplitude now becomes a function of $z$ and we represent it as ${\cal A}_n(z)$. The original amplitude is simply given as:
\begin{align}
	{\cal A}_n(0)&=\oint \frac{dz}{z}~{\cal A}_n(z)
\end{align}
where the contour is a circle around origin. If we now restrict our attention to tree-level diagrams, then ${\cal A}_n(z)$ has no branch cuts and only has simple poles (denoted by $z_{\alpha}$) that occur when the propagator goes onshell. Deforming the above contour and noting that when the propagator goes onshell, the amplitude separates into a product of two lower point functions, we obtain the BCFW recursion relations. 

While deforming the contour, apart from picking up all the simple poles $z_{\alpha}$, we also pick a contribution as $z\rightarrow \infty $. Ideally, the BCFW formalism is most convenient, when this boundary term vanishes. When we use BCFW relations for four point amplitude of scalar Compton scattering by shifting the momenta of photons (see \cite{Cheung} for discussion of BCFW in six dimensions for massless amplitudes),  we find that the contribution at $z\rightarrow \infty $ does not vanish . That is, the ``naive'' shift that was used in the massless case  does not work in the massive case. There are some works in the literature (see \cite{BCFW boudary} for example) which discuss how to proceed with BCFW recursion relations if one has ``boundary'' terms at infinity and we leave such endeavors for future work.

\section*{Acknowledgments}

We thank Avinash Raju for early collaboration and emphasizing to us that $SO(5) \sim Sp(4)$.

\appendix

\section{Four Dimensions}
There are some inconsequential (but potentially confusing) typos in \cite{Nima}, so we have solved and reproduced here the results of massive spinor helicity formalism in $4D$ in different conventions. We have also calculated some of the results from a different perspective using Lorentz boost. 
\subsection{Results in the Convention $x^{\mu}=(t, -\vec{x})$ as in arXiv:1709.04891}

The conventions adopted in \cite{Nima} are as follows (natural units are employed everywhere): 
\begin{itemize}
\item $g_{\mu \nu} = (+, -, -, -)$
\item $x^{\mu}=(t, -\vec{x}) \Rightarrow p_{\mu}=(E, \vec{p})$\footnote{For maximal compatibility with the results in \cite{Nima}, we find that we need this convention. Results in the standard convention $x^{\mu}=(t, \vec{x})$ are dealt in sub-appendix \ref{standardconvention}.}
\item $\epsilon_{ij} = \left(
\begin{array}{cc}
 0 & -1 \\
 1 & 0 \\
\end{array}
\right) \Rightarrow \epsilon^{ij}= \left(
\begin{array}{cc}
 0 & 1 \\
 -1 & 0 \\
\end{array}
\right)$. This is the convention adopted for both $\epsilon_{\alpha \dot{\alpha}}$ and $\epsilon_{IJ}$ where $\alpha$ and $\dot{\alpha}$ are Lorentz indices while $I$ and $J$ are little group indices, used as raising/lowering operators.
\item $4-$ vector Pauli matrices are $\sigma^{\mu} = (1, \vec{\sigma})$ and $\bar{\sigma}^{\mu} = (1, -\vec{\sigma})$ where $\sigma^1 = \left(
\begin{array}{cc}
 0 & 1 \\
 1 & 0 \\
\end{array}
\right)$, $\sigma^2 = \left(
\begin{array}{cc}
 0 & -i \\
 i & 0 \\
\end{array}
\right)$ and $\sigma^3 = \left(
\begin{array}{cc}
 1 & 0 \\
 0 & -1 \\
\end{array}
\right)$.

\end{itemize}

Corresponding to these conventions, we have the momentum bispinor matrix as follows:
\begin{equation}
p_{\alpha \dot{\alpha}} = \sigma.p = (\sigma^{\mu})_{\alpha \dot{\alpha}} p_{\mu} = \left(
\begin{array}{cc}
 E + p \cos(\theta) &  p \sin (\theta ) e^{-i \phi } \\
  p \sin (\theta ) e^{i \phi } & E - p \cos(\theta) \\
\end{array}
\right)
\end{equation}
and
\begin{equation}
p^{\dot{\alpha} \alpha} = \bar{\sigma}.p = (\bar{\sigma}^{\mu})^{\dot{\alpha} \alpha} p_{\mu} = \left(
\begin{array}{cc}
 E - p \cos(\theta) &  -p \sin (\theta ) e^{-i \phi } \\
  -p \sin (\theta ) e^{i \phi } & E + p \cos(\theta) \\
\end{array}
\right)
\end{equation}

\subsubsection{Massless Case}

The Lorentz group in this case is Spin$(3,1) \cong SL(2, \mathbb{C})$ and the little group is $SO(2) \cong U(1)$. The massless spinors $\lambda_{\alpha}$ and $\tilde{\lambda}_{\dot{\alpha}}$ satisfy the following equations:
\begin{equation}
p_{\alpha \dot{\alpha}} = \lambda_{\alpha} \tilde{\lambda}_{\dot{\alpha}}
\end{equation}

The forms of $\lambda$ and $\tilde{\lambda}$ that satisfy this equation are as follows:
\begin{equation}
\lambda_{\alpha} = \sqrt{2E} \left(
\begin{array}{cc}
 c \\
  s \\
\end{array}
\right), \hspace{7mm} \tilde{\lambda}_{\dot{\alpha}} = \sqrt{2E} \left(
\begin{array}{cc}
 c \\
  s^{*} \\
\end{array}
\right)
\end{equation}

Here $c \coloneqq \cos(\frac{\theta}{2})$, $s \coloneqq \sin(\frac{\theta}{2}) e^{i \phi}$ and $s^{*} \coloneqq \sin(\frac{\theta}{2}) e^{-i \phi}$. They happen to satisfy the following massless Dirac equations (or Weyl equations):
\begin{equation}
p_{\alpha \dot{\alpha}} \lambda^{\alpha} = 0,  \hspace{7mm} p_{\alpha \dot{\alpha}} \tilde{\lambda}^{\dot{\alpha}} = 0
\end{equation}

\subsubsection{Massive Case}

In the massive case, the Lorentz group remains the same but the little group changes to $SO(3) \cong SU(2)$. The latter contributes an index to both the spinors $\lambda_{\alpha}$ and $\tilde{\lambda}_{\dot{\alpha}}$. The spinors satisfy the following equation:
\begin{equation}
p_{\alpha \dot{\alpha}} = \lambda_{\alpha}^I \tilde{\lambda}_{\dot{\alpha} I}
\end{equation}

This is a little group invariant quantity as little group index is summed over. This has to be true by the very definition of the little group. The forms of $\lambda_{\alpha}^I$ and $\tilde{\lambda}_{\dot{\alpha}}^I$ that satisfy this equation are as follows\footnote{We believe that the expression for $\tilde{\lambda}_{\dot{\alpha}}^I$ has a typo in \cite{Nima} (eq. $(C.2)$) following which a systematic error is carried forward. Their calculations of scattering amplitudes remain unaffected throughout their paper because the explicit forms are not required anywhere in those calculations.}:
\begin{equation}
\lambda_{\alpha}^I = \left(
\begin{array}{cc}
 \sqrt{E + p} c & - \sqrt{E - p} s^{*} \\
 \sqrt{E + p} s & \sqrt{E - p} c \\
\end{array}
\right), \hspace{7mm} \tilde{\lambda}_{\dot{\alpha}}^I = \left(
\begin{array}{cc}
 \sqrt{E-p} s & \sqrt{E+p} c \\
 - \sqrt{E - p} c & \sqrt{E+p} s^{*} \\
\end{array}
\right)
\end{equation}

Determinants of both of these spinors are equal to $m$. They happen to satisfy the following massive Dirac equations:
\begin{equation}
p_{\alpha \dot{\alpha}} \tilde{\lambda}^{\dot{\alpha} I} = - m \lambda_{\alpha}^I, \hspace{7mm} p_{\alpha \dot{\alpha}} \lambda^{\alpha I} = m \tilde{\lambda}_{\dot{\alpha}}^I
\end{equation}

Now to evaluate the high energy limit, it is preferable to decompose the aforementioned $\lambda_{\alpha}^I$ and $\tilde{\lambda}_{\dot{\alpha}}^I$ as follows:
\begin{equation}
\label{l}
\lambda_{\alpha}^I = \sqrt{E+p} \zeta_{\alpha}^{+} \zeta^{-I} + \sqrt{E-p} \zeta_{\alpha}^{-} \zeta^{+I} 
\end{equation}
and
\begin{equation}
\label{ltil}
\tilde{\lambda}_{\dot{\alpha}}^I = \sqrt{E+p} \tilde{\zeta}_{\dot{\alpha}}^{-} \zeta^{+I} + \sqrt{E-p} \tilde{\zeta}_{\dot{\alpha}}^{+} \zeta^{-I}
\end{equation}
where $\zeta^{\mp I}$ are the basis vectors spanning two dimensional spinor space, chosen to be $\zeta^{-I}=\left(
\begin{array}{cc}
 1 & 0 \\
\end{array}
\right)$ and $\zeta^{+I}=\left(
\begin{array}{cc}
 0 & 1 \\
\end{array}
\right)$ so that they satisfy the normalization condition $\epsilon_{IJ} \zeta^{+I} \zeta^{-J} = 1$. With this choice of basis vectors, we have the following expressions for $\zeta_{\alpha}^{\pm}$ and $\tilde{\zeta}_{\dot{\alpha}}^{\mp}$:
\begin{equation}
\label{zeta}
\zeta_{\alpha}^{+} = \left(
\begin{array}{cc}
 c \\
 s \\
\end{array}
\right), \tilde{\zeta}_{\dot{\alpha}}^{-} = \left(
\begin{array}{cc}
 c \\
 s^{*} \\
\end{array}
\right), \zeta_{\alpha}^{-} = \left(
\begin{array}{cc}
 - s^{*} \\
 c \\
\end{array}
\right), \tilde{\zeta}_{\dot{\alpha}}^{+} = \left(
\begin{array}{cc}
 s \\
 - c \\
\end{array}
\right)
\end{equation}

Here we define $\lambda_{\alpha} \coloneqq \sqrt{E+p} \zeta_{\alpha}^{+}$, $\eta_{\alpha} \coloneqq \sqrt{E-p} \zeta_{\alpha}^{-}$, $\tilde{\lambda}_{\dot{\alpha}} \coloneqq \sqrt{E+p} \tilde{\zeta}_{\dot{\alpha}}^{-}$ and $\tilde{\eta}_{\dot{\alpha}} \coloneqq \sqrt{E-p} \tilde{\zeta}_{\dot{\alpha}}^{+}$. This enables us to write equations ({\ref{l}}) and ({\ref{ltil}}) as follows:
\begin{equation}
\lambda_{\alpha}^I = \lambda_{\alpha} \zeta^{-I} + \eta_{\alpha} \zeta^{+I} 
\end{equation}
and
\begin{equation}
\tilde{\lambda}_{\dot{\alpha}}^I = \tilde{\lambda}_{\dot{\alpha}} \zeta^{+I} + \tilde{\eta}_{\dot{\alpha}} \zeta^{-I}
\end{equation}

Making identifications with angle and square spinor notations $\lambda_{\alpha} \rightarrow |p \rangle_{\alpha} \Rightarrow \lambda^{\alpha} \rightarrow \langle p|^{\alpha}$ and $\tilde{\lambda}_{\dot{\alpha}} \rightarrow [p|_{\dot{\alpha}} \Rightarrow \tilde{\lambda}^{\dot{\alpha}} \rightarrow |p]^{\dot{\alpha}}$, we obtain the following results:
\begin{equation}
\label{ln}
\langle \lambda \eta\rangle = \lambda^{\alpha} \eta_{\alpha} = \epsilon^{\alpha \beta} \lambda_{\beta} \eta_{\alpha} = -m = - \langle \eta \lambda \rangle
\end{equation}
and
\begin{equation}
\label{lntil}
[\tilde{\lambda} \tilde{\eta}] = \tilde{\lambda}_{\dot{\alpha}} \tilde{\eta}^{\dot{\alpha}} = \tilde{\lambda}_{\dot{\alpha}} \epsilon^{\dot{\alpha} \dot{\beta}} \tilde{\eta}_{\dot{\beta}} = -m = -[\tilde{\eta} \tilde{\lambda}]
\end{equation}

Rewriting $\eta_{\alpha} = \frac{\sqrt{E+p}}{\sqrt{E+p}} \sqrt{E-p} \zeta_{\alpha}^{-} = m \hat{\eta}_{\alpha}$ where $\hat{\eta}_{\alpha} = \frac{\zeta_{\alpha}^{-}}{\sqrt{E+p}}$ and similarly $\tilde{\eta}_{\dot{\alpha}} = m \hat{\tilde{\eta}}_{\dot{\alpha}}$ where $\hat{\tilde{\eta}}_{\dot{\alpha}} = \frac{\tilde{\zeta}_{\dot{\alpha}}^{+}}{\sqrt{E+p}}$, equations ({\ref{ln}}) and ({\ref{lntil}}) become the following{\footnote{In order to make $\langle \lambda \eta \rangle = [\tilde{\lambda} \tilde{\eta}] = m \Rightarrow \langle \lambda \hat{\eta} \rangle = [\tilde{\lambda} \hat{\tilde{\eta}}] = 1$, we must chose basis vectors $\zeta^{\mp I}$ such that $\epsilon_{IJ} \zeta^{+I} \zeta^{-J} = -1$, for example $\zeta^{- I}= \left(
\begin{array}{cc}
 1 & 0 \\
\end{array}
\right)$ and $\zeta^{+I}=\left(
\begin{array}{cc}
 0 & -1 \\
\end{array}
\right)$. With this choice of basis vectors, eq. ({\ref{zeta}}) gets modified accordingly as $\tilde{\zeta}_{\dot{\alpha}}^{-} = \left(
\begin{array}{cc}
 - c \\
 - s^{*} \\
\end{array}
\right), \zeta_{\alpha}^{-} = \left(
\begin{array}{cc}
 s^{*} \\
 - c \\
\end{array}
\right)$ while $\zeta_{\alpha}^{+}$ and $\tilde{\zeta}_{\dot{\alpha}}^{+}$ remain unchanged. Equations ({\ref{l}}) and (\ref{ltil}) continue to hold by making these adjustments.}:
\begin{equation}
\langle \lambda \hat{\eta} \rangle = -1, \hspace{7mm} [\tilde{\lambda} \hat{\tilde{\eta}}] = - 1
\end{equation}

\subsection{Results in the Standard Convention $x^{\mu}=(t, +\vec{x})$}
\label{standardconvention}

The conventions adopted in all our calculations are as follows (again, natural units are employed everywhere): 
\begin{itemize}
\item $g_{\mu \nu} = (+, -, -, -)$
\item $x^{\mu}=(t, +\vec{x}) \Rightarrow p_{\mu}=(E, -\vec{p})$
\item $\epsilon_{ij} = \left(
\begin{array}{cc}
 0 & -1 \\
 1 & 0 \\
\end{array}
\right) \Rightarrow \epsilon^{ij}= \left(
\begin{array}{cc}
 0 & 1 \\
 -1 & 0 \\
\end{array}
\right)$. This is the convention adopted for both $\epsilon_{\alpha \dot{\alpha}}$ and $\epsilon_{IJ}$ where $\alpha$ and $\dot{\alpha}$ are Lorentz indices while $I$ and $J$ are little group indices, used as raising/lowering operators.
\item $4-$ vector Pauli matrices are $\sigma^{\mu} = (1, \vec{\sigma})$ and $\bar{\sigma}^{\mu} = (1, -\vec{\sigma})$ where $\sigma^1 = \left(
\begin{array}{cc}
 0 & 1 \\
 1 & 0 \\
\end{array}
\right)$, $\sigma^2 = \left(
\begin{array}{cc}
 0 & -i \\
 i & 0 \\
\end{array}
\right)$ and $\sigma^3 = \left(
\begin{array}{cc}
 1 & 0 \\
 0 & -1 \\
\end{array}
\right)$.

\end{itemize}

These are the standard convention that is found in literature. In this convention we get the momentum bispinor matrix as follows:
\begin{equation}
p_{\alpha \dot{\alpha}} = \sigma.p = (\sigma^{\mu})_{\alpha \dot{\alpha}} p_{\mu} = \left(
\begin{array}{cc}
 E - p \cos(\theta) &  -p \sin (\theta ) e^{-i \phi } \\
  -p \sin (\theta ) e^{i \phi } & E + p \cos(\theta) \\
\end{array}
\right)
\end{equation}
and
\begin{equation}
p^{\dot{\alpha} \alpha} = \bar{\sigma}.p =  (\bar{\sigma}^{\mu})^{\dot{\alpha} \alpha} p_{\mu} = \left(
\begin{array}{cc}
 E + p \cos(\theta) &  p \sin (\theta ) e^{-i \phi } \\
  p \sin (\theta ) e^{i \phi } & E - p \cos(\theta) \\
\end{array}
\right)
\end{equation}

\subsubsection{Massless Case}
Again we have $p_{\alpha \dot{\alpha}} = \lambda_{\alpha} \tilde{\lambda}_{\dot{\alpha}}$. The forms of $\lambda$ and $\tilde{\lambda}$ that satisfy this equation are as follows:
\begin{equation}
\lambda_{\alpha} = \sqrt{2E} \left(
\begin{array}{cc}
 - s^{*} \\
  c \\
\end{array}
\right), \hspace{7mm} \tilde{\lambda}_{\dot{\alpha}} = \sqrt{2E} \left(
\begin{array}{cc}
 -s \\
  c \\
\end{array}
\right)
\end{equation}

They happen to satisfy the following massless Dirac equations (or Weyl equations):
\begin{equation}
p_{\alpha \dot{\alpha}} \lambda^{\alpha} = 0,  \hspace{7mm} p_{\alpha \dot{\alpha}} \tilde{\lambda}^{\dot{\alpha}} = 0
\end{equation}

\subsubsection{Massive Case}

Again we have $p_{\alpha \dot{\alpha}} = \lambda_{\alpha}^I \tilde{\lambda}_{\dot{\alpha} I}$. The forms of $\lambda_{\alpha}^I$ and $\tilde{\lambda}_{\dot{\alpha}}^I$ that satisfy this equation are as follows:
\begin{equation}
\lambda_{\alpha}^I = \left(
\begin{array}{cc}
 -\sqrt{E + p} s^{*} & - \sqrt{E - p} c \\
 \sqrt{E + p} c & -\sqrt{E - p} s \\
\end{array}
\right), \hspace{7mm} \tilde{\lambda}_{\dot{\alpha}}^I = \left(
\begin{array}{cc}
  \sqrt{E-p} c & - \sqrt{E+p} s \\
 \sqrt{E - p} s^{*} & \sqrt{E+p} c \\
\end{array}
\right)
\end{equation}

Determinants of both of these spinors are equal to $m$. They happen to satisfy the following massive Dirac equations:
\begin{equation}
p_{\alpha \dot{\alpha}} \tilde{\lambda}^{\dot{\alpha} I} = - m \lambda_{\alpha}^I, \hspace{7mm} p_{\alpha \dot{\alpha}} \lambda^{\alpha I} = m \tilde{\lambda}_{\dot{\alpha}}^I
\end{equation}

Now to evaluate the high energy limit, it is preferable to decompose the aforementioned $\lambda_{\alpha}^I$ and $\tilde{\lambda}_{\dot{\alpha}}^I$ as follows:
\begin{equation}
\label{l2}
\lambda_{\alpha}^I = \sqrt{E+p} \zeta_{\alpha}^{+} \zeta^{-I} + \sqrt{E-p} \zeta_{\alpha}^{-} \zeta^{+I} 
\end{equation}
and
\begin{equation}
\label{ltil2}
\tilde{\lambda}_{\dot{\alpha}}^I = \sqrt{E+p} \tilde{\zeta}_{\dot{\alpha}}^{-} \zeta^{+I} + \sqrt{E-p} \tilde{\zeta}_{\dot{\alpha}}^{+} \zeta^{-I}
\end{equation}
where $\zeta^{\mp I}$ are the basis vectors spanning two dimensional spinor space, chosen to be $\zeta^{-I}=\left(
\begin{array}{cc}
 1 & 0 \\
\end{array}
\right)$ and $\zeta^{+I}=\left(
\begin{array}{cc}
 0 & -1 \\
\end{array}
\right)$ so that they satisfy the normalization condition $\epsilon_{IJ} \zeta^{+I} \zeta^{-J} = -1$. With this choice of basis vectors, we have the following expressions for $\zeta_{\alpha}^{\pm}$ and $\tilde{\zeta}_{\dot{\alpha}}^{\mp}$:
\begin{equation}
\label{zeta2}
\zeta_{\alpha}^{+} = \left(
\begin{array}{cc}
 -s^{*} \\
 c \\
\end{array}
\right), \tilde{\zeta}_{\dot{\alpha}}^{-} = \left(
\begin{array}{cc}
 s \\
 - c \\
\end{array}
\right), \zeta_{\alpha}^{-} = \left(
\begin{array}{cc}
 c \\
 s \\
\end{array}
\right), \tilde{\zeta}_{\dot{\alpha}}^{+} = \left(
\begin{array}{cc}
 c \\
 s^{*} \\
\end{array}
\right)
\end{equation}

Here we define $\lambda_{\alpha} \coloneqq \sqrt{E+p} \zeta_{\alpha}^{+}$, $\eta_{\alpha} \coloneqq \sqrt{E-p} \zeta_{\alpha}^{-}$, $\tilde{\lambda}_{\dot{\alpha}} \coloneqq \sqrt{E+p} \tilde{\zeta}_{\dot{\alpha}}^{-}$ and $\tilde{\eta}_{\dot{\alpha}} \coloneqq \sqrt{E-p} \tilde{\zeta}_{\dot{\alpha}}^{+}$. This enables us to write equations ({\ref{l2}}) and ({\ref{ltil2}}) as follows:
\begin{equation}
\lambda_{\alpha}^I = \lambda_{\alpha} \zeta^{-I} + \eta_{\alpha} \zeta^{+I} 
\end{equation}
and
\begin{equation}
\tilde{\lambda}_{\dot{\alpha}}^I = \tilde{\lambda}_{\dot{\alpha}} \zeta^{+I} + \tilde{\eta}_{\dot{\alpha}} \zeta^{-I}
\end{equation}

Making identifications with angle and square spinor notations $\lambda_{\alpha} \rightarrow |p \rangle_{\alpha} \Rightarrow \lambda^{\alpha} \rightarrow \langle p|^{\alpha}$ and $\tilde{\lambda}_{\dot{\alpha}} \rightarrow [p|_{\dot{\alpha}} \Rightarrow \tilde{\lambda}^{\dot{\alpha}} \rightarrow |p]^{\dot{\alpha}}$, we obtain the following results:
\begin{equation}
\label{ln2}
\langle \lambda \eta \rangle = \lambda^{\alpha} \eta_{\alpha} = \epsilon^{\alpha \beta} \lambda_{\beta} \eta_{\alpha} = +m = - \langle \eta \lambda \rangle
\end{equation}
and
\begin{equation}
\label{lntil2}
[\tilde{\lambda} \tilde{\eta}] = \tilde{\lambda}_{\dot{\alpha}} \tilde{\eta}^{\dot{\alpha}} = \tilde{\lambda}_{\dot{\alpha}} \epsilon^{\dot{\alpha} \dot{\beta}} \tilde{\eta}_{\dot{\beta}} = +m = -[\tilde{\eta} \tilde{\lambda}]
\end{equation}

Rewriting $\eta_{\alpha} = \frac{\sqrt{E+p}}{\sqrt{E+p}} \sqrt{E-p} \zeta_{\alpha}^{-} = m \hat{\eta}_{\alpha}$ where $\hat{\eta}_{\alpha} = \frac{\zeta_{\alpha}^{-}}{\sqrt{E+p}}$ and similarly $\tilde{\eta}_{\dot{\alpha}} = m \hat{\tilde{\eta}}_{\dot{\alpha}}$ where $\hat{\tilde{\eta}}_{\dot{\alpha}} = \frac{\tilde{\zeta}_{\dot{\alpha}}^{+}}{\sqrt{E+p}}$, equations ({\ref{ln2}}) and ({\ref{lntil2}}) become the following:
\begin{equation}
\langle \lambda \hat{\eta} \rangle = +1 = - \langle \hat{\eta} \lambda \rangle, \hspace{7mm} [\tilde{\lambda} \hat{\tilde{\eta}}] = +1 = - [\hat{\tilde{\eta}} \tilde{\lambda}]
\end{equation}

\subsection{Obtaining Massive Spinors via Lorentz Boost}


We can also construct massive helicity spinors in the following way using Lorentz boost but we are not going to use this approach and the corresponding explicit results elsewhere in the paper.

Consider a moving Dirac spinor in 'chiral' basis $\Psi = (\phi_{R}(v), \chi_{L}(v))^{T}$ where $v$ is the speed, that takes the form $\phi_{R}(0) = \chi_{L}(0)$ (which is true for massive Dirac spinor at rest) and transforms as the following:
\begin{equation}
\Psi \rightarrow \left(
\begin{array}{cc}
 \Lambda(v) & 0 \\
 0 & \Lambda(-v) \\
\end{array}
\right) \Psi
\end{equation}

This is a $4 \times 4$ matrix where $\Lambda(v)$ and $\Lambda(-v)$ are defined as follows:
\begin{equation}
\Lambda(v) = \exp \left(- \frac{\vec{\sigma} . \vec{\rho}}{2} + i \frac{\vec{\sigma} . \vec{\theta}}{2} \right)
\end{equation} 
and
\begin{equation}
\Lambda(-v) =(\Lambda(v)^{\dagger})^{-1}= \exp \left( \frac{\vec{\sigma} . \vec{\rho}}{2} + i \frac{\vec{\sigma} . \vec{\theta}}{2} \right)
\end{equation}

In the spinor space, the rest frame momentum bispinor is given by the matrix $M=\left(
\begin{array}{cc}
 m & 0 \\
 0 & m \\
\end{array}
\right)$. Boosting this gives the momentum bispinor having a general momentum $\vec{p}$ and energy $E$ as shown in the following:
\begin{equation}
M \rightarrow N M N^{\dagger}
\end{equation}
where
\begin{equation}
N = \sqrt{\frac{E+m}{2m}} \left( 1 - \frac{\vec{\sigma}.\vec{p}}{E+m} \right) = \sqrt{\frac{E+m}{2m}} \left(
\begin{array}{cc}
 \left(1-\frac{p \cos (\theta )}{E+m}\right) & -\frac{ p \sin (\theta ) e^{-i \phi }}{ (E+m)} \\
 -\frac{ p \sin (\theta ) e^{i \phi }}{(E+m)} & \left( 1 + \frac{p \cos (\theta )}{E+m} \right) \\
\end{array}
\right)
\end{equation}
$N$ and $N^{\dagger}$ have determinants equal to $1$ as expected since $N$ belongs to $SL(2, \mathbb{C)}$ group.

This gives what we call $p_{\alpha \dot{\alpha}}$ as follows:
\begin{equation}
\label{paa}
p_{\alpha \dot{\alpha}} = N M N^{\dagger} = \left(
\begin{array}{cc}
 E - p \cos(\theta) & - p \sin (\theta ) e^{-i \phi } \\
 - p \sin (\theta ) e^{i \phi } & E + p \cos(\theta) \\
\end{array}
\right)
\end{equation}

Similarly to obtain $p^{\dot{\alpha} \alpha}$, we use $M \rightarrow N^{'} M N^{' \dagger}$ where $N^{'}$ is given by the following:
\begin{equation}
N^{'} = \sqrt{\frac{E+m}{2m}} \left( 1 + \frac{\vec{\sigma}.\vec{p}}{E+m} \right) = \sqrt{\frac{E+m}{2m}} \left(
\begin{array}{cc}
 \left(1+\frac{p \cos (\theta )}{E+m}\right) & \frac{ p \sin (\theta ) e^{-i \phi }}{ (E+m)} \\
 \frac{ p \sin (\theta ) e^{i \phi }}{(E+m)} & \left( 1 - \frac{p \cos (\theta )}{E+m} \right) \\
\end{array}
\right)
\end{equation}
$N^{'}$ and $N^{' \dagger}$ have determinants equal to $1$ as expected since $N^{'}$ belongs to $SL(2, \mathbb{C)}$ group. 

This gives the following result upon boosting from the rest frame:
\begin{equation}
p^{\dot{\alpha} \alpha} = N^{'} M N^{' \dagger} = \left(
\begin{array}{cc}
 E + p \cos(\theta) &  p \sin (\theta ) e^{-i \phi } \\
  p \sin (\theta ) e^{i \phi } & E - p \cos(\theta) \\
\end{array}
\right)
\end{equation}

Note that the conventions used by us are: $g_{\mu \nu} = (+, -, -, -)$, $\epsilon_{12}=-1$ where $\epsilon_{\alpha \beta} / \epsilon^{\alpha \beta}$ and $\epsilon_{\dot{\alpha} \dot{\beta}} / \epsilon^{\dot{\alpha} \dot{\beta}}$ are the lowering/raising operators and $x^{\mu} = (t, \vec{x}) \Rightarrow p^{\mu} = (E, \vec{p}) \Rightarrow p_{\mu} = (E, -\vec{p})$. 

We can write the rank $2$ matrix (eq. ({\ref{paa}})) as $p_{\alpha \dot{\alpha}} = \lambda_{\alpha}^I \tilde{\lambda}_{\dot{\alpha} \hspace{0.5mm} I}$. Writing the rest frame momentum bispinor matrix as $p_{o, \alpha \dot{\alpha}} = p_{o}^{\dot{\alpha} {\alpha}} = \left(
\begin{array}{cc}
 m &  0 \\
  0 & m \\
\end{array}
\right) = \left(\begin{array}{cc} \sqrt{m} \\ 0 \\ \end{array} \right) \left(\begin{array}{cc} \sqrt{m} & 0 \\ \end{array} \right) + \left(\begin{array}{cc} 0 \\ \sqrt{m} \\ \end{array} \right) \left(\begin{array}{cc} 0 & \sqrt{m} \\ \end{array} \right)$, we get the following results for $\lambda_{\alpha}^I$ and $\tilde{\lambda}_{\dot{\alpha}}^I$ obtained via boosting through $N$ and $N^{'}$ respectively:
\begin{equation}
\lambda_{\alpha}^I = \frac{1}{\sqrt{2(E + m)}} \left(
\begin{array}{cc}
 E + m - p \cos(\theta) & - p \sin (\theta ) e^{-i \phi } \\
 - p \sin (\theta ) e^{i \phi } & E + m + p \cos(\theta) \\
\end{array}
\right)
\end{equation}
and 
\begin{equation}
\tilde{\lambda}_{\dot{\alpha}}^I = \frac{1}{\sqrt{2(E + m)}} \left(
\begin{array}{cc}
p \sin (\theta ) e^{i \phi } & E + m - p \cos(\theta)  \\
 -(E + m + p \cos(\theta)) &  - p \sin (\theta ) e^{-i \phi } \\
\end{array}
\right)
\end{equation}

These satisfy:
\begin{itemize}
\item[i.] $p_{\alpha \dot{\alpha}} = \lambda_{\alpha}^I \tilde{\lambda}_{\dot{\alpha} \hspace{0.5mm} I}$
\item[ii.] $p_{\alpha \dot{\alpha}} \tilde{\lambda}^{\dot{\alpha} \hspace{0.5mm} I} = - m \lambda_{\alpha}^I$
\item[iii.] $p_{\alpha \dot{\alpha}} \lambda^{\alpha \hspace{0.5mm} I} = m \tilde{\lambda}_{\dot{\alpha}}^I$
\item[iv.] $det(\lambda) = \det(\tilde{\lambda}) = m$
\end{itemize}

\section{Six Dimensions}
\subsection{Conventions}
The conventions adopted are as follows:
\begin{itemize}
\item $g_{\mu \nu} = (+, -, -, -)$
\item $x^{\mu}=(t, +\vec{x}) \Rightarrow p_{\mu}=(E, -\vec{p})$
\item $\epsilon_{ij} = \left(
\begin{array}{cc}
 0 & -1 \\
 1 & 0 \\
\end{array}
\right) \Rightarrow \epsilon^{ij}= \left(
\begin{array}{cc}
 0 & 1 \\
 -1 & 0 \\
\end{array}
\right)$. This is the convention adopted for both $\epsilon_{\dot{\alpha} \dot{\beta}}$ and $\epsilon_{\alpha \beta}$ where $\dot{\alpha}$ and $\alpha$ belongs to the first and the second $SU(2)$ respectively of the $6D$ massless little group $SO(4) \cong SU(2) \times SU(2)$.
\item The invariant tensor of the group $Sp(4)$ is taken to be $J_{IJ}= \left(\begin{array}{cc} 0_{2\times 2} & 1_{2\times 2} \\ -1_{2\times 2} & 0_{2\times 2} \end{array} \right)$.
\item $4-$ vector Pauli matrices are $\sigma^{\mu} = (1, \vec{\sigma})$ and $\bar{\sigma}^{\mu} = (1, -\vec{\sigma})$ where $\sigma^1 = \left(
\begin{array}{cc}
 0 & 1 \\
 1 & 0 \\
\end{array}
\right)$, $\sigma^2 = \left(
\begin{array}{cc}
 0 & -i \\
 i & 0 \\
\end{array}
\right)$ and $\sigma^3 = \left(
\begin{array}{cc}
 1 & 0 \\
 0 & -1 \\
\end{array}
\right)$.
\item $6D-$Pauli matrices are of the form $\Sigma^{\mu}= (\gamma ^{0}, \gamma ^{1}, \gamma ^{2}, \gamma ^{3}, \gamma ^{4}, \gamma ^{5})$ and $\bar{\Sigma}^{\mu} = (\tilde{\gamma}^{0}, \tilde{\gamma}^{1}, \tilde{\gamma}^{2}, \tilde{\gamma}^{3}, \tilde{\gamma}^{4}, \tilde{\gamma}^{5})$. They are taken to be as follows:
\begin{itemize}
\item[a.] $\gamma ^{0} = i \sigma^{1} \otimes \sigma^{2}$, $\tilde{\gamma}^{0} = -i \sigma^{1} \otimes \sigma^{2}$
\item[b.] $\gamma ^{1} = i \sigma^{2} \otimes \sigma^{3}$, $\tilde{\gamma}^{1} = i \sigma^{2} \otimes \sigma^{3}$
\item[c.] $\gamma ^{2} = - \sigma^{2} \otimes \sigma^{0}$, $\tilde{\gamma}^{2} = \sigma^{2} \otimes \sigma^{0}$
\item[d.] $\gamma ^{3} = - i \sigma^{2} \otimes \sigma^{1}$, $\tilde{\gamma}^{3} = -i \sigma^{2} \otimes \sigma^{1}$
\item[e.] $\gamma ^{4}= - \sigma^{3} \otimes \sigma^{2}$, $\tilde{\gamma}^{4} = \sigma^{3} \otimes \sigma^{2}$
\item[f.] $\gamma ^{5} = i \sigma^{0} \otimes \sigma^{2}$, $\tilde{\gamma}^{5} = i \sigma^{0} \otimes \sigma^{2}$
\end{itemize}
\end{itemize}

In this convention we get the momentum bispinor matrix as follows:
\begin{equation}
p_{A B} = \Sigma.p = (\Sigma^{\mu})_{A B} p_{\mu} = \left(
\begin{array}{cccc}
 0 & -i p_4-p_5 & -p_1-i p_2 & E+p_3 \\
 i p_4+p_5 & 0 & p_3-E & p_1-i p_2 \\
 p_1+i p_2 & E-p_3 & 0 & i p_4-p_5 \\
 -E-p_3 & i p_2-p_1 & p_5-i p_4 & 0 \\
\end{array}
\right)
\end{equation}
and
\begin{equation}
p^{AB} = \bar{\Sigma}.p =  (\bar{\Sigma}^{\mu})^{AB} p_{\mu} = \left(
\begin{array}{cccc}
 0 & i p_4-p_5 & -p_1+i p_2 & p_3-E \\
 -i p_4+p_5 & 0 & p_3+E & p_1+i p_2 \\
 p_1-i p_2 & -E-p_3 & 0 & -i p_4-p_5 \\
 E-p_3 & -p_1 - i p_2 & p_5+i p_4 & 0 \\
\end{array}
\right)
\end{equation}
where $\vec{p} = (p_1, p_2, p_3, p_4, p_5)$ is used.

\subsection{Angle/Square Brackets and Lorentz Invariant Objects in $6D$}

We introduce the angle and square brackets notation using the following equations:
\begin{equation}
\label{angsqmom}
p_{AB} = \tilde{\lambda}_A^I \tilde{\lambda}_{B I} = |p]_A^I [p|_{B I}, \hspace{7mm} p^{AB} = \lambda^A_I \lambda^{B I} = |p \rangle^A_I \langle p|^{B I}
\end{equation}

As we can see, the same spinor solutions are used to construct each type of momentum matrix, hence the following results must hold:
\begin{equation}
\label{angsq}
[p|_{A I} J^{I J} = [p|_A^J = |p]_A^J, \hspace{7mm} |p \rangle^A_I J^{I J} = |p \rangle^{A J} = \langle p|^{A J}
\end{equation}
where $J^{I J}$ is the inverse of $J_{I J}$ which is the tensor left invariant under $Sp(4)$ group transformation.

Let particles be labelled by lower-case Latin indices. With all the aforementioned conventions, various Lorentz invariant quantities can be constructed that transform covariantly under the little group transformation. They are as follows:
\begin{itemize}
\item[i.] $\lambda_i^{A I} \tilde{\lambda}_{j A}^J = \tilde{\lambda}_{j A}^I \lambda_i^{A J} = \langle i|j]$ where $\langle i|j]$ has little group indices as $\langle i^I|j^J]$.

\item[ii.] $\lambda_{i I}^{A} \tilde{\lambda}_{j A J} = \tilde{\lambda}_{j A I} \lambda_{i J}^{A} = [j|i \rangle $ but due to (\ref{angsq}), this is essentially same as $(i)$ above. Here too the little group indices are $[j_J|i_I \rangle$.

\item[iii.] $\epsilon_{ABCD} \lambda_i^{A I} \lambda_j^{B J} \lambda_k^{C K} \lambda_l^{D L} = \langle i^I j^J k^K l^L \rangle$ that can be simply written as $\langle i j k l \rangle$.

\item[iv.] $\epsilon^{ABCD} \tilde{\lambda}_{i A I} \tilde{\lambda}_{j B J} \tilde{\lambda}_{k C K} \tilde{\lambda}_{l D L} = [i_I j_J k_K l_L]$ that can be simply written as $[i j k l]$.

\end{itemize}

Using (\ref{angsqmom}) we see that the mass dimensions of each angle and square spinor is $\sqrt{m}$. Constructing the scalar product $p.q$ using these and $Tr[\Sigma^{\mu} \bar{\Sigma}^{\nu}] = Tr[\bar{\Sigma}^{\mu} \Sigma^{\nu}] = 4 \eta^{\mu \nu}$, we get the following results:
\begin{equation}
\label{4p.q}
4 p.q = 4 \eta^{\mu \nu} p_{\mu} q_{\nu} = \langle q^{A J}|p_A^I] [p_{B I}| q^B_J \rangle
\end{equation}

The momentum conservation $\sum_{i}p_i = 0$ takes the following form using (\ref{angsqmom}):
\begin{equation}
\label{momcons1}
\sum_i \langle q | i] [i|k \rangle = 0
\end{equation}
or
\begin{equation}
\label{momcons2}
\sum_i [q|i \rangle \langle i | k] = 0
\end{equation}
where $q$ and $k$ are reference spinors. Eqs. (\ref{momcons1}) and (\ref{momcons2}) follow from first and second equations of (\ref{angsqmom}) respectively.

The explicit form of $[p| q \rangle$ calculated using the spinor solutions obtained earlier is as follows:
\begin{equation}
[p | q \rangle = m J_{IJ} \hspace{3.5mm} \Rightarrow \hspace{3.5mm} Det([p | q \rangle ) = (m^2)^2
\end{equation}

In massless case, determinant becomes $0$ thereby rendering $[p|q \rangle$ to be rank $1$ matrix whereas in the massive case here, the rank is $2$. So there is no problem in taking inverses of these Lorentz invariant, little group covariant objects in the massive case unlike the massless case.

\subsection{$p^{AB} = \lambda^{AI} \lambda^{B}_{I}$}

Solving the massless Dirac equation $p_{AB} \lambda^{A a} = 0$, where $a$ is the second $SU(2)$ little group index of the $SO(4) \cong SU(2) \times SU(2)$, gives the massless spinor solution $\lambda^{A a}$ in the limit $E \rightarrow p$ where $p$ denotes the norm of $\vec{p}$. Similarly another spinor solution $\eta^{A \dot{a}}$ is obtained by taking the massless limit $E \rightarrow -p$. The latter (unphysical) massless limiting condition is important in order to systematically construct the massive spinor as shown in eq. (\ref{massl}). Both of these solutions correspond to zero eigenvalue of the corresponding momentum bispinor matrix. They satisfy the following equations:
\begin{equation}
p^{AB} = \lambda^{Aa} \lambda^{B}_{a}, p'^{AB} = \eta^{A \dot{a}} \eta^{B}_{\dot{a}}
\end{equation}
where $\vec{p} \rightarrow -\vec{p}$ in $p'^{AB}$. The forms of both the spinors are as follows:
\begin{equation}
\label{masslessl}
\lambda^A_a =\frac{1}{ \sqrt{-i p_4 - p_5}}\left(
\begin{array}{cc}
 p_1- i p_2 & p_3 - p \\
 -(p + p_3) & p_1 + i p_2  \\
 0 & - i p_4 - p_5 \\
 - i p_4 - p_5 & 0 \\
\end{array}
\right),
\eta^{A \dot{a}} = \frac{1}{ \sqrt{i p_4 + p_5}}\left(
\begin{array}{cc}
 -(p + p_3) & p_1 - i p_2  \\
 - (p_1 + i p_2)  & p - p_3  \\
 i p_4 + p_5 & 0 \\
 0 & -(i p_4 + p_5) \\
\end{array}
\right)
\end{equation}

The massive case in $D=6$ has the same Lorentz group $SU^{*}(4)$ but the little group changes to $SO(5) \cong Sp(4)$. Then the massive spinor $\lambda^{A I}$ (index $I$ runs over the group $Sp(4)$) is constructed using the above calculated massless spinors as follows:
\begin{equation}
\label{massl}
\lambda^{A I} = \sqrt{\frac{E+p}{2 p}} \lambda^A_a e^{+I a} + \sqrt{\frac{E-p}{2p}} \eta^{A \dot{a}} e^{-I}_{\dot{a}}
\end{equation}
where $e^{+I a}$ and $ e^{-I}_{\dot{a}}$ are the basis vectors in the spinor space explicitly given as follows:
\begin{equation}
\label{eia}
e^{+I a} = \left(\begin{array}{cccc} 0 & +i & 0 & 0 \\ 0 & 0 & 0 & +i \end{array} \right)
\end{equation}
and
\begin{equation}
\label{e-ia}
e^{-I}_{\dot{a}} = \left(\begin{array}{cccc} -1 & 0 & 0 & 0 \\ 0 & 0 & +1 & 0 \\ \end{array} \right)
\end{equation}
where the basis vectors satisfy the following normalization conditions:
\begin{equation}
\label{normal}
e^{+I a} J_{IJ} e^{+J b} = e^{-I}_{\dot{a}} J_{IJ} e^{-J}_{\dot{b}} = \left( \begin{array}{cc} 0 & -1 \\ +1 & 0 \\ \end{array} \right), e^{+I a} J_{IJ} e^{-J}_{\dot{b}} = e^{-I}_{\dot{a}} J_{IJ} e^{+J b} = 0
\end{equation}

The coefficients in eq. (\ref{massl}) are chosen in order to satisfy the determinant condition $det(\lambda^{AI})=m^2$ which is natural consequence of the equation $p^{AB} = \lambda^{AI} \lambda^{A}_I$ where $det(p^{AB})=(m^2)^2$. Also the coefficients conveniently ensure that the massless condition in the limit $E \rightarrow p$ is recovered smoothly. Only the first term survives and the coefficient goes to $1$. 

Plugging eqs. (\ref{masslessl}), (\ref{eia}) and (\ref{e-ia}) in eq. (\ref{massl}) give the following massive spinor form:
\begin{equation}
\label{massivel}
\lambda^{A}_{I} = \frac{1}{\sqrt{-i p_4 - p_5}}\left(
\begin{array}{cccc}
 \sqrt{\frac{E - p}{2 p}} (ip_1+ p_2) &  -i\sqrt{\frac{E + p}{2 p}} (p- p_3) & -i\sqrt{\frac{ E - p}{2 p}} (p+ p_3) & -\sqrt{\frac{E + p}{2p}} ( ip_1 +p_2) \\
 i\sqrt{\frac{E- p}{2 p}} (p- p_3) &  \sqrt{\frac{E +p}{2 p}} (ip_1 -  p_2) & \sqrt{\frac{E - p}{2 p}} (p_2 - i p_1) & i \sqrt{\frac{E + p}{2 p}} (p+ p_3) \\
 0 & i \sqrt{\frac{E + p}{2 p}} ( p_4 - ip_5) & \sqrt{\frac{E - p}{2 p}} ( -p_4 + ip_5) & 0 \\
 \sqrt{\frac{E - p}{2 p}} ( p_4 -i p_5) & 0 & 0 & -i \sqrt{\frac{E+p}{2 p}} ( p_5 - ip_4) \\
\end{array}
\right)
\end{equation}

This satisfies the following equations:
\begin{itemize}
\item[i.] $Det(\lambda^{A I}) = (E - p) (E + p) = m^2$

\item[ii.] $p^{AB} = \lambda^{AI} \lambda^{B}_{I}$. Note that the indices are written in the same order as the massless case, namely $p^{AB} = \lambda^{Aa} \lambda^{B}_{a}$.

\item[iii.] Massless limit $E \rightarrow p$ matches with the known massless spinor helicity formalism where the spinor satisfies the massless Dirac equation $p_{AB} \lambda^A_a = 0$ or $p_{AB} \lambda^{A a} = 0$.

\item[iv.] Actions of momentum bispinor acting on massless spinors are: 
\begin{equation}
\label{plambda}
p_{AB} \lambda^A_a = \left(E-p \right) \tilde{\eta}_{B a}, \hspace{7mm} p_{AB} \eta^{A \dot{a}} = \left( E+p \right) \tilde{\lambda}_{B}^{\dot{a}}
\end{equation}

\end{itemize}

\subsection{$p_{AB} = \tilde{\lambda}_{AI} \tilde{\lambda}_{B}^{I}$}

We can write massive spinor $\tilde{\lambda}_A^I$ in terms of massless spinors $\tilde{\lambda}_A^{\dot{a}}$ and $\tilde{\eta}_{A a}$ as follows:
\begin{equation}
\label{tildeml}
\tilde{\lambda}_A^I = \sqrt{\frac{E+p}{2 p}} \tilde{\lambda}_A^{\dot{a}} e^{-I}_{\dot{a}} + \sqrt{\frac{E-p}{2 p}} \tilde{\eta}_{A a} e^{+I a}
\end{equation}

Here $\dot{a}$ belongs to the first $SU(2)$ of the massless little group $SO(4) \cong SU(2) \times SU(2)$ and $I$ belongs to the massive little group $Sp(4)$ which is congruent to $SO(5)$. The massless spinors are obtained by solving the massless Dirac equation $p^{AB} \tilde{\lambda}_{A}^{\dot{a}} = 0$ and $p'^{AB} \tilde{\eta}_{A a} = 0$ where $p'^{AB}$ has $\vec{p} \rightarrow -\vec{p}$. Both these spinor solutions correspond to zero eigenvalue of the corresponding momentum bispinor matrices. They satisfy $p_{AB} = \tilde{\lambda}_{A \dot{a}} \tilde{\lambda}_{A}^{\dot{a}}$ and $p'_{AB} = \tilde{\eta}_{Aa} \tilde{\eta}_{A}^{a}$ where again $p'_{AB}$ means that the substitution $\vec{p} \rightarrow -\vec{p}$ is made in $p_{AB}$. Again, the unphysical massless limiting condition $E\rightarrow -p$ is required to systematically construct the massive spinor as clear in eq. (\ref{tildeml}).

The basis vectors in the spinor space, $e^{-I}_{\dot{a}}$ and $e^{+I a}$, are the same objects as used in eq. (\ref{massl}). Their explicit forms are exactly the same as in eqs. (\ref{e-ia}) and (\ref{eia}) respectively. This makes the normalization of the basis vectors same as in eq. (\ref{normal}).

The coefficients in eq. (\ref{tildeml}) are decided by the condition $det(\tilde{\lambda}_A^I) = m^2$ which is a natural consequence of the relation $p_{AB} = \tilde{\lambda}_{AI} \tilde{\lambda}_{A}^{I}$ where $det(p_{AB}) = (m^2)^2$. Also the coefficients ensure that the massless limit is recovered smoothly when $E\rightarrow p$. In this limit, the second term dies and the coefficient of first term goes to $1$, thereby leaving only $\tilde{\lambda}_A^{\dot{a}}$ which was obtained in the first place by solving the massless Dirac equation in the limit $E\rightarrow p$. 

Plugging everything in eq. (\ref{tildeml}) gives the massive spinor $\tilde{\lambda}_A^I$ as follows:
\begin{equation}
\label{massivetildel}
\tilde{\lambda}_{AI} = \frac{1}{\sqrt{-i p_4 - p_5}} \left(
\begin{array}{cccc}
 \sqrt{\frac{E+p}{2 p}} (-p_4 + i p_5) & 0 & 0 &  \sqrt{\frac{E-p}{2 p}} (p_4 - i p_5) \\
 0 &  \sqrt{\frac{E-p}{2 p}} (p_4- i p_5) & \sqrt{\frac{E+p}{2 p}} (-p_4 + i p_5) & 0 \\
 -i \sqrt{\frac{E+p}{2 p}} (p- p_3) & \sqrt{\frac{E-p}{2 p}} (p_2 - i p_1) & i \sqrt{\frac{E+p}{2 p}} (p_1+ i p_2) & -i \sqrt{\frac{E-p}{2 p}} (p+ p_3) \\
 i\sqrt{\frac{E+p}{2 p}} (p_1 - i p_2) & - i \sqrt{\frac{E-p}{2 p}} (p- p_3) & - i \sqrt{\frac{E+p}{2 p}} (p+ p_3) & -\sqrt{\frac{E-p}{2 p}} (i p_1 + p_2) \\
\end{array}
\right)
\end{equation}

This satisfies the following equation:
\begin{itemize}
\item[i.] $Det(\tilde{\lambda}_A^I) = (E-p)(E+p) = m^2$

\item[ii.] $p_{AB} = \tilde{\lambda}_{AI} \tilde{\lambda}_{B}^{I}$. Note that the indices are written in the same order as the massless case, namely $p_{AB} = \tilde{\lambda}_{A \dot{a}} \tilde{\lambda}_{B}^{\dot{a}}$.

\item[iii.] Massless limit $E \rightarrow p$ matches with the known massless spinor helicity formalism satisfying the massless Dirac equation $p^{AB} \tilde{\lambda}_{A \dot{a}} = 0$ or $p^{AB} \tilde{\lambda}_{A}^{\dot{a}} = 0$.

\item[iv.] Actions of momentum bispinor acting on massless spinors are: 
\begin{equation}
\label{plambdatilde}
p^{AB} \tilde{\lambda}_A^{\dot{a}} = \left( E-p \right) \eta^{B \dot{a}}, \hspace{7mm} p^{AB} \tilde{\eta}_{A a} = \left( E+p \right) \lambda^{B}_{a}
\end{equation}

\end{itemize}

\subsection{Massive Dirac Equation}

The spinors $\lambda^{A I}$ as constructed in eq. (\ref{massl}) and explicitly shown in eq. (\ref{massivel}) as well as $\tilde{\lambda}_A^I$ in eqs. (\ref{tildeml}) and (\ref{massivetildel}) satisfy the following massive Dirac equation in $6D$:
\begin{equation}
\label{massiveDirac}
p_{AB} \lambda^{A I} = m \tilde{\lambda}_B^I, \hspace{7mm} p^{AB} \tilde{\lambda}_A^I = m \lambda^{B I}
\end{equation}

\section{Summary of Important Conventions/Definitions in $6D$}

To summarize, we are using the following conventions/definitions:
\begin{gather}
p^{AB} p_{BC}=m^2\delta ^A_C ; \qquad p_{AB} p^{BC}=m^2\delta_A ^C  \\
p_{AB}\lambda ^{ AI}=m\tilde{\lambda }_B^I; \qquad p^{AB}\tilde{\lambda } _{A}^I=m\lambda ^{BI}  \\
p^{AB}=\lambda ^{AI}\lambda ^{B}_I; \qquad p_{AB}=\tilde{\lambda }_{AI}\tilde{\lambda }_{B}^I  \\
\lambda ^{A}_I \tilde{\lambda }_{AJ} = mJ_{IJ}; \qquad \lambda ^{AI}\tilde{\lambda } _{A}^J=-mJ^{IJ}  \\
\lambda ^{AI}\tilde{\lambda } _{AJ} = m\delta ^I_J; \qquad \lambda ^{A}_I\tilde{\lambda } _{A}^J=-m\delta ^J_I  \\
\lambda ^{AI}\tilde{\lambda } _{BI} = m\delta _B^A; \qquad \lambda ^{A}_I\tilde{\lambda } _{B}^I=-m\delta _B^A  \\
J_{IJ}\lambda ^{AJ} = \lambda ^A_I; \qquad J^{IJ}\lambda ^A_J=\lambda ^{AI} \\
J_{IJ}J^{JK} = \delta _I^K  \\
e^{-I}_{\dot{a}}J_{IJ}e^{-J}_{\dot{b}} = \epsilon_{\dot{a} \dot{b}}; \qquad e^{+Ia}J_{IJ}e^{+Jb}=-\epsilon ^{{a}{b}} \\
p_{AB}\lambda ^{Aa} =(E-p)\tilde{\eta }_B^a ; \qquad p_{AB}\eta ^A_{\dot{a}}=(E+p)\tilde{\lambda }_{B\dot{a}}  \\
p^{AB}\tilde{\eta }_A^a = (E+p)\lambda ^{Ba}; \qquad p^{AB}\tilde{\lambda }_{A\dot{a}}=(E-p)\eta ^B_{\dot{a}} \\
\lambda^{Ab} \epsilon_{ab} = \lambda^A_a; \qquad \tilde{\lambda}_{A\dot{b}} \epsilon^{\dot{a} \dot{b}}=\tilde{\lambda}_{A}^{\dot{a}} \\
\lambda ^{Aa}\tilde{\eta }_{A}^b = -2p \epsilon^{{a}{b}}; \qquad  \eta^A_{\dot{a}}\tilde{\lambda }_{A\dot{b}}=-2p \epsilon_{\dot{a} \dot{b}} \\ 
p^{AB} =\frac{1}{2}\epsilon^{ABCD} p_{CD} \\
p_{AB}q^{AB} = -4p.q  \\
[p_{\dot{b}}q_e\rangle ^{-1}\langle q_ep_{\dot{a}}] =\delta _{\dot{a}}^{\dot{b}}; \qquad \langle q_bp_{\dot{a}}]^{-1}[p_{\dot{a}}q_c\rangle =\delta _c^b  \\
\langle p_aq^{\dot{b}}]^{-1} = -\frac{\langle p^aq_{\dot{b}}]}{2p.q} \\
 \langle p^a|q|p^b\rangle = -2\epsilon^{{a}{b}} (p.q); \qquad  [p^{\dot{a}}|q|p^{\dot{b}}]=+2\epsilon^{\dot{a} \dot{b}}(p.q) 
\end{gather}
The identity involving Levi-Civita tensor of $SU(4)$ are as follows:
\begin{align}
\epsilon_{ABEF}\epsilon ^{CDGH}=\left|
\begin{array}{cccc} 
\delta^C_A & \delta^C_B & \delta^C_E & \delta^C_F \\ 
\delta^D_A & \delta^D_B & \delta^D_E & \delta^D_F \\
\delta^G_A & \delta^G_B & \delta^G_E & \delta^G_F \\
\delta^H_A & \delta^H_B & \delta^H_E & \delta^H_F \\ 
\end{array}
\right| 
\end{align}
\begin{align}
= \delta ^C_A\left[\delta ^D_B\left(\delta ^G_E\delta ^H_F-\delta ^H_E\delta ^G_F\right)-\delta ^D_E\left(\delta ^G_B\delta ^H_F-\delta ^H_B\delta ^G_F\right)+\delta ^D_F\left(\delta ^G_B\delta ^H_E-\delta ^H_B\delta ^G_E\right)\right]\nonumber \\
-\delta ^C_B\left[\delta ^D_A\left(\delta ^G_E\delta ^H_F-\delta ^H_E\delta ^G_F\right)-\delta ^D_E\left(\delta ^G_A\delta ^H_F-\delta ^H_A\delta ^G_F\right)+\delta ^D_F\left(\delta ^G_A\delta ^H_E-\delta ^H_A\delta ^G_E\right)\right]\nonumber \\
+\delta ^C_E\left[\delta ^D_A\left(\delta ^G_B\delta ^H_F-\delta ^H_B\delta ^G_F\right)-\delta ^D_B\left(\delta ^G_A\delta ^H_F-\delta ^H_A\delta ^G_F\right)+\delta ^D_F\left(\delta ^G_A\delta ^H_B-\delta ^H_A\delta ^G_B\right)\right]\nonumber \\
-\delta ^C_F\left[\delta ^D_A\left(\delta ^G_B\delta ^H_E-\delta ^H_B\delta ^G_E\right)-\delta ^D_B\left(\delta ^G_A\delta ^H_E-\delta ^H_A\delta ^G_E\right)+\delta ^D_E\left(\delta ^G_A\delta ^H_B-\delta ^H_A\delta ^G_B\right)\right]
\end{align}


\begin{thebibliography}{99}


\bibitem{TASI} 
C.~Cheung,
``TASI Lectures on Scattering Amplitudes,''
arXiv:1708.03872 [hep-ph].

\bibitem{Nima} 
N.~Arkani-Hamed, T.~C.~Huang and Y.~t.~Huang, ``Scattering Amplitudes For All Masses and Spins,'' arXiv:1709.04891 [hep-th].



\bibitem{Elvang}
H.~Elvang and Y.~t.~Huang, ``Scattering Amplitudes,'' arXiv:1308.1697 [hep-th].


\bibitem{all dimensions} 
R.~H.~Boels and D.~O'Connell,
``Simple superamplitudes in higher dimensions,''
JHEP {\bf 1206}, 163 (2012)
[arXiv:1201.2653 [hep-th]].

\bibitem{Cheung} 
C.~Cheung and D.~O'Connell, ``Amplitudes and Spinor-Helicity in Six Dimensions,'' JHEP {\bf 0907}, 075 (2009) 
[arXiv:0902.0981 [hep-th]].


\bibitem{Georgi}
H.~Georgi, ``Lie Algebras In Particle Physics. From Isospin To Unified Theories,'' Front.\ Phys.\  {\bf 54}, 1 (1982).

\bibitem{Plefka} 
J.~Plefka, T.~Schuster and V.~Verschinin,
``From Six to Four and More: Massless and Massive Maximal Super Yang-Mills Amplitudes in $6D$ and $4D$ and their Hidden Symmetries,'' JHEP {\bf 1501}, 098 (2015)
[arXiv:1405.7248 [hep-th]].

\bibitem{KLT}
H.~Kawai, D.~C.~Lewellen and S.~H.~H.~Tye, ``A Relation Between Tree Amplitudes of Closed and Open Strings,'' Nucl.\ Phys.\ B {\bf 269}, 1 (1986).

\bibitem{bcfw1}
R.~Britto, F.~Cachazo, B.~Feng and E.~Witten, ``Direct proof of tree-level recursion relation in Yang-Mills theory,'' Phys.\ Rev.\ Lett.\  {\bf 94}, 181602 (2005)
[arXiv:hep-th/0501052].


\bibitem{Badger}
S.~D.~Badger, E.~W.~N.~Glover, V.~V.~Khoze and P.~Svrcek, ``Recursion relations for gauge theory amplitudes with massive particles,'' JHEP {\bf 0507}, 025 (2005)
[arXiv:hep-th/0504159].

\bibitem{6d to 4d loops} 
Z.~Bern, J.~J.~Carrasco, T.~Dennen, Y.~t.~Huang and H.~Ita,
``Generalized Unitarity and Six-Dimensional Helicity,''
Phys.\ Rev.\ D {\bf 83}, 085022 (2011)
[arXiv:1010.0494 [hep-th]].


\bibitem{DHS} 
T.~Dennen, Y.~t.~Huang and W.~Siegel,
``Supertwistor space for 6D maximal super Yang-Mills,''
JHEP {\bf 1004}, 127 (2010)
[arXiv:0910.2688 [hep-th]].



\bibitem{BCFW boudary} 
B.~Feng, J.~Wang, Y.~Wang and Z.~Zhang,
``BCFW Recursion Relation with Nonzero Boundary Contribution,''
JHEP {\bf 1001}, 019 (2010)
[arXiv:0911.0301 [hep-th]].


\end{thebibliography}
\end{document}